# Enhanced Power Point Tracking for High Hysteresis Perovskite Solar Cells: A Galvanostatic Approach

Emilio J. Juarez-Perez[1,2]*, Cristina Momblona[1], Roberto Casas[3], Marta Haro[1,4]

1: Nanostructured Films & Particles Research Group (NFP). Instituto de Nanociencia y Materiales de Aragón (INMA). CSIC-Universidad de Zaragoza, Zaragoza 50009, Spain.

2: Aragonese Foundation for Research and Development (ARAID) Government of Aragon. Zaragoza 50018, Spain

3: Howlab - Human Openware Research Group., Aragon Institute of Engineering Research (I3A), University of Zaragoza, 50018 Zaragoza, Spain

4: Departamento de Química Física. Facultad de Ciencias. Universidad de Zaragoza. Zaragoza 50009, Spain

Corresponding Author: ejjuarezperez@unizar.es

## Abstract

This article introduces a novel Maximum Power Point Tracking (MPPT) algorithm and cost-effective hardware for long-term operational stability measurements in perovskite solar cells (PSCs). Harnessing the untapped potential of solar energy sources is crucial for achieving a sustainable future, and accurate MPPT is vital to maximizing power generation. However, existing MPPT algorithms for classical photovoltaic technology lead to suboptimal performance and decreased energy efficiency conversion when applied to the most stable perovskite devices, the so-called triple mesoscopic hole transport material (HTM)-free metal halide PSCs. To address this challenge, our research focuses on developing an innovative low-cost hardware solution for research purposes that enables massive long-term stability measurements, eliminating the need for expensive and complex stability monitoring systems. Our galvanostatic MPPT algorithm ensures continuous and precise tracking achieving superior operational performance for high hysteresis PSCs. The suggested enhancements bear significant implications for the extensive integration of perovskite solar cell technologies, particularly those dependent on power optimizer devices.

# 1. Introduction

Electricity generated by solar photovoltaic (PV) systems has unique advantages compared to other renewable energies (wind, hydroelectricity, biomass, among others) with a lower environmental impact and expanded integration possibilities in future society and mobility applications. As an example, silicon panels are already integrated into the urban and rural areas because of their low-cost installation and low maintenance, obtaining energy independence with long-term savings. In recent decades, novel PV technologies have emerged, with perovskite solar cells (PSCs) standing out as the most promising technology with power conversion efficiency (PCE) record comparable to that of single-crystal silicon cells.[1] However, the potential industrial relevancy of halide perovskite technology is yet compromised as the assurance of long-term operational stability has not yet been guaranteed. Analyzing and ensuring the operational stability of PSCs is a complex endeavor, contingent upon numerous factors and demanding significant economic and time investments. To promote and accelerate research and market adoption of this technology, especially given their stability challenges and hysteresis phenomena, it is crucial to have accessible and specialized equipment for widespread and high throughput operational stability measurements in PSCs.

The factors affecting the stability of PSCs are extrinsic –temperature, humidity, $O_2$– , or intrinsic by the choice of perovskite chemical composition, charge transport materials, metal electrodes and device layout.[2-9] In recent years, PSC utilizing carbon electrodes have emerged as a notably stable device architecture of choice.[10] Solar cells using metallic top electrodes – gold or silver – often suffer from stability issues such as oxidation of the electrode and metal migration to the perovskite layer inducing degradation in the device.[11] Furthermore, carbon-based PSCs obviate the need for hole transport materials (HTMs), thereby mitigating potential stability concerns stemming from the incorporation of dopants to enhance layer conductivity or the absence of adverse side reactions with halide perovskite materials.[12,13] Nonetheless, it is crucial to acknowledge that while this heightened operational stability is achieved, there can be a trade-off involving a decrease in device efficiency when compared to HTM-based counterparts. HTM-free carbon-based PSCs especially present hysteresis phenomenon and complicate the accurate evaluation of the cell performance. Hysteresis in PSCs is observed by the difference between current density-voltage (JV) curves upon a change in voltage sweep direction. The hysteresis in PSCs is not an intrinsic characteristic of the device but dependent on the voltage scan parameters selected for the realization of the JV curve[14,15] and there is no agreement on its origin.[16-22] Within the context of this work, hysteresis in the JV curve characteristics introduces notable uncertainty in the determination of the maximum power point (the cell voltage producing the maximum power output, $V_{MPP}$) and consequently, the PCE of the device. To address this challenge, researchers envisaged a stabilized output power (SOP) measurement setting the device at constant $V_{MPP}$, previously determined through the JV curve, for a short interval of minutes to ensure a genuinely stable power output.[23,24] In recent years of perovskite solar cell research, a crucial shift has occurred in the efficiency measurement methodology, with a growing emphasis on this kind of steady-state efficiency assessments at fixed voltages to ensure the credibility of the reported results. There is a growing consensus that the traditional Maximum Power Point Tracking (MPPT) algorithms, commonly used to maximize power output under variable irradiation of well-established PV technologies,[25-28] should assume greater significance as the standard methodology for performance assessment in PSCs instead of the traditional fixed voltage rate JV curve measurements.[24,29-31] MPPT algorithms offer a realistic assessment of device efficiency in real-world scenarios and provide insights in the short-term operational stability. Early efforts in developing MPPT algorithms for PSCs were based on the Perturb and Observe (P&O) algorithm, revealing that devices with significant hysteresis required modeling of current data decay to accurate tracking[32] or minimizing large oscillations around optimal voltage by introducing the concept of power thresholds before changing voltage direction in the algorithm.[33] Other variations of the P&O algorithm reported include using two/three voltage point measurements or short JV sweeps[34] and stabilization times (10 s, 1 s).[35,36] Additionally, other algorithms known to be used in PSCs include the Genetic Algorithm[37] and the fractional open-circuit voltage tracking a circuit model of a PSC under variable illumination conditions.[38] Certainly, despite MPPT benefits in PSC research, limited number of potentiostats and solar simulators equipment in the labs, along with challenges in algorithm implementation, hinder widespread use of this technique for statistically significant batch testing in emerging photovoltaic labs.

Evolving computing -memory and operations per second- capacities of microelectronics together with its capacity to acquire and generate signal at high speed, allow to implement complex algorithms in tiny devices at very low cost. This work presents the development of a cost-effective open hardware-based platform[39] for long-term stability measurements on lab-scale solar cells eliminating the need for expensive monitoring systems. Accurate tracking of the MPPT is crucial for maximizing power generation. However, existing MPPT firmware algorithms designed for classical photovoltaic technology perform suboptimal in metal halide perovskite-based single cells due to hysteresis. To address this, a novel MPPT algorithm is implemented here which controlling current (galvanostatic approach) instead of voltage bias (potentiostatic approach) applied in the device enables continuous and precise tracking of the maximum power output resulting in superior operational performance. This article outlines the detailed methodology employed to construct such custom hardware and implements useful algorithms for short operational stability tests on small area lab-scale single cell PSCs. These advancements have significant implications for widespread adoption of perovskite solar cell technologies in solar energy harvesting, driving progress towards a greener and more sustainable future.

## 2. Methods and Experimental Section.
### 2.1 Perovskite Solar Cell Assembly and encapsulation.
The assembly of the PSCs was carried out using the commercial triple mesoscopic monolithic Perovskite Solar Cell Kit provided by Solaronix. This kit offered a ready-to-use perovskite precursor solution and monolithic electrodes containing compact $TiO_2$, mesoporous $TiO_2$, mesoporous $ZrO_2$, and carbon layer, all optimized for their respective thicknesses. The precursor solution was carefully dropped onto the electrode stack substrate, allowing it to permeate the triple layer mesoscopic porous structure. To ensure precise and controlled device fabrication, we applied an adhesive impregnation mask to prevent the spread of perovskite precursor solution beyond the active area. Through the annealing process (1.5 h, 50ºC), the perovskite material grew within the mesoporous triple stack followed by a humidity assisted thermal exposure (HTE) treatment.[40] The humidity treatment was performed in an oven at 40 °C containing a vessel with saturated NaCl aqueous solution to reach constant 75% of relative humidity.[41] The temperature and relative humidity inside the oven were continuously monitored by a sensor module (DHT22, Adafruit). Then, devices were enclosed using the standard pre-laminated glass lids with sealing gaskets from the kit as primary encapsulation stage. Copper cables were soldered to metal spring clamps attached to the electrodes to establish the electrical connections. A mask of 0.64 cm$^2$ was placed to delimit the active area of the solar cell. To monitor the device temperature, a 10KΩ - 25°C NTC type thermistor was fixed in contact to the back surface of the PSC ensuring that the thermistor head is not exposed to direct light illumination. Finally, the assembled device was placed in a specific jig and completely immersed in two-component transparent epoxy resin followed by a 48-hour curing process working as a secondary encapsulation stage (Figure 1.a.b).

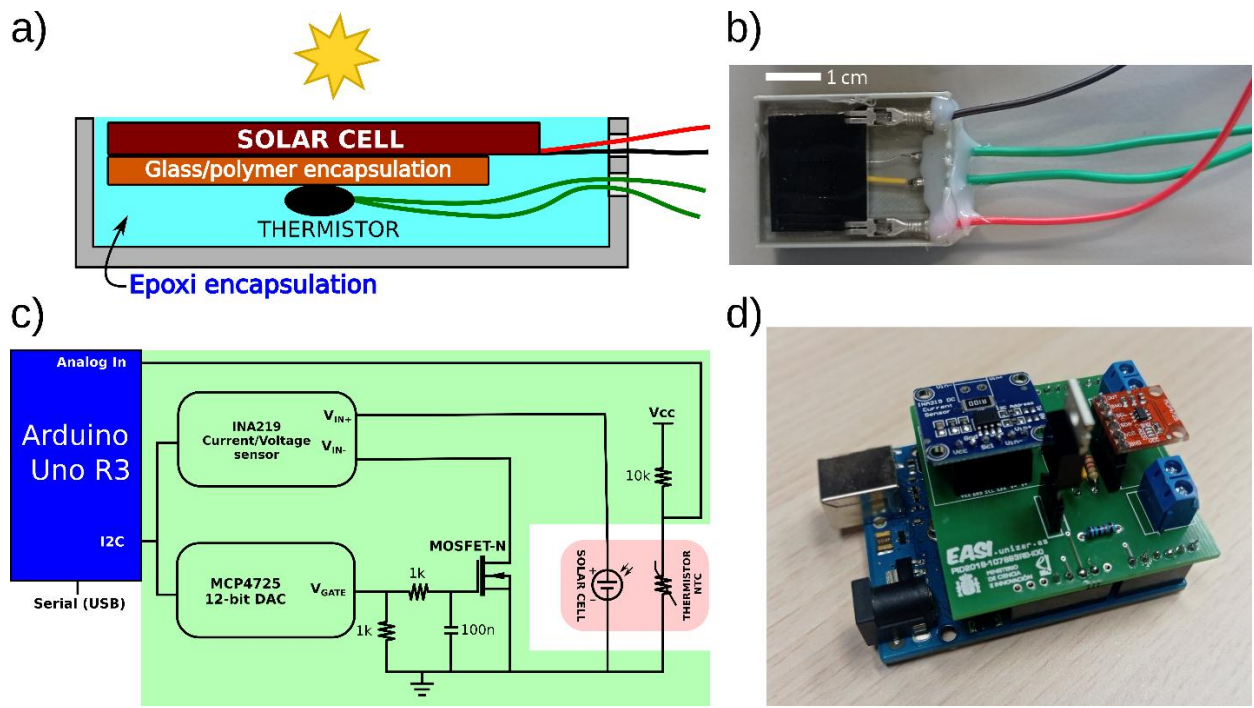

**Figure 1.** a) Schematic cross-section of the assembled solar cell device, b) top view of the physical device. c) Schematic representation of the Arduino board connected to INA219 and MCP4725 breakout boards, N-MOSFET, solar cell, thermistor, and USB serial connection to the PC. The pink zone highlights the encapsulated components referred in panel (b) and the green zone highlights components in the Arduino Shield. d) Image of the Arduino device connected to the Shield module containing the INA219 and MCP4725 breakouts, N-MOSFET and screw terminal block connectors for solar cell and thermistor.

**2.2 JV curve and operational stability measurement setup.**
The JV measurements were performed under simulated sunlight illumination using a LED solar simulator (LSH-7320, Newport, AM1.5G, 100 mW/cm²). A commercial white LED was also eventually employed as a light source by matching the equivalent photocurrent in the device to 1 Sun of simulated illumination power, which is explicitly stated in the JV or MPPT measurement.

For the potentiostatic type measurement, a source-meter unit (X200, Ossila) was used connected in series with the device. The JV characteristics of the PSC and a commercial monocrystal silicon single solar cell (KXOB22-12X1, IXYS, 1.18 cm$^2$), acting as a reference of hysteresis-free device, were acquired through voltage sweeps from -0.1 V to 1.2 V and -0.1 V to 0.65 V, respectively. The voltage step used an increment of 10 mV and a stabilization time of 0.005 s. The galvanostatic JV measurement of the solar cells were conducted using the custom hardware developed as part of this study and detailed in the following section.

Independently of the potentiostatic or galvanostatic method used, a JV curve was recorded where $J_{SC}$, $V_{OC}$, FF and PCE of the solar cells were calculated using the formula: PCE (%) = ($J_{SC} \cdot V_{OC} \cdot FF$)·100 / $P_{IN}$, where $J_{SC}$ is the short-circuit current density, $V_{OC}$ is the open-circuit voltage, FF = ($V_{MPP} \cdot J_{MPP}$)/($V_{OC} \cdot J_{SC}$) is the fill factor and $P_{IN}$ is the incident power density. Voltage at the maximum power output ($V_{MPP}$) and current density at the maximum power output ($J_{MPP}$) were determined as the voltage and current corresponding to the maximum power output ($P_{MAX} = V_{MPP} \cdot J_{MPP}$) achieved by the photovoltaic device, respectively. Series ($R_{SERIES}$) and shunt resistance ($R_{SHUNT}$) were extracted from the JV curve fit using the single diode model in Grapa.[42]

Photovoltaic operational stability measurements of the device under potentiostatic or galvanostatic conditions were conducted using the Perturb & Observe (P&O) MPPT algorithm in two different modes of operation. The first mode involved alternating full JV scans to ascertain the $V_{MPP}$ (or $J_{MPP}$), followed by a stage of Stabilized Output Power (SOP) at $V_{MPP}$ over a defined time interval. Multiple consecutive JV-SOP cycles were performed to ensure consistency and stability of the solar cell. The purpose of this first mode is using MPPT techniques for performance evaluation of PSCs under constant illumination (Section 3.3). The second mode involves the conventional implementation of the Perturb and Observe (P&O) MPPT algorithm for outdoor performance under varying irradiation conditions (Section 3.5). It employs a fixed step voltage increment for perturbation. In this work, the device using the second mode P&O algorithm underwent a variable irradiation sequence test in compliance with the EN 50530 standard.[43,44] This test protocol comprised a single cycle, delineated into four distinct phases. Firstly, there is a gradual increase in irradiance, transitioning from 30 mW/cm$^2$ to 100 mW/cm$^2$ over a duration of 300 seconds. Subsequently, the irradiance level is maintained at a constant 100 mW/cm$^2$ for the following 300 seconds. Next, a decline in irradiance is implemented, returning to 30 mW/cm$^2$, again spanning a 300-second interval. Finally, the irradiance level remains fixed at 30 mW/cm$^2$ for an additional 300 seconds.

## 2.3 Hardware and Firmware Design

The hardware design of the system incorporates a hierarchical architecture comprising various components, each serving specific functions. The main components include an 8-bit microcontroller (ATmega328); a 12-bit digital-to-analog converter (MCP4725) that generates an analog voltage that bias the gate terminal of a N-channel MOSFET transistor (IRLZ34N, 68 W max. of power dissipation); a 12-bit digital power monitor (INA219) that measures voltage and current of the photovoltaic device; and an NTC 10kOhm thermistor (MF52) for monitoring the working temperature of the device. The microcontroller serves as the central processing unit driving the solar cell. Communication with this microcontroller is facilitated by a conventional computer, allowing firmware algorithm upload, data retrieval, storage and data processing. This interaction is facilitated through a USB-based serial connection and external components, utilizing both digital and analog input/output pins, as depicted in Figure 1.c.

The N-MOSFET plays a pivotal role acting as a variable electronic load driving the photovoltaic device in a galvanostatic manner because no controlled bias is applied in the photovoltaic device terminals.[45] In the typical setup, the photovoltaic device was connected in series with the drain and source terminals of the N-MOSFET which is operated in its ohmic region by applying a continuous voltage in the gate terminal ($V_{GATE}$) by the DAC which is commanded by the microcontroller enabling a variable load in the solar cell terminals.

For firmware development and microcontroller upload, the Arduino Integrated Development Environment (IDE) was utilized. This IDE offers a user-friendly interface for code writing, compiling, and uploading to the Arduino microcontroller. The firmware code was written in the Arduino programming language, which is based on the C/C++ programming language, enabling efficient control and operation of the system. Python scripts were developed for collecting USB serial data from the Arduino to the computer and performing postprocessing for graphical representation (matplotlib). The firmware in this work implements the first and second mode operation of the P&O MPPT algorithms for the microcontroller. However, using an open-hardware/software platform, users can modify the microcontroller's firmware to run custom algorithms for short-term stability assessment, such as the asymptotic $P_{MAX}$ scan,[31] dynamical IVs[30] or more advanced transient MPPT sequences.[29]

A frozen version of all codes used in this study is provided in the Support Information (SI) and they will be publicly available on a GitHub[46] repository encouraging further development and contributions from the interested research community. SI file provides in-depth details about the selection process for the appropriate TTL level N-MOSFET considering specific requirements and criteria for the Arduino platform and a calibration procedure for the INA219 sensor. Circuit diagrams were created using computer-aided design free software (Fritzing) and the design deposited as SI. The circuits were then checked in a prototyping breadboard and then a custom PCB shield was created for the Arduino. Gerber files are deposited as SI files.

The codename of the device (Arduino + shield) is "Perovskino" and the algorithms implemented are "floating voltage" or "passive/galvanostatic approach" type methods.

## 3. Results and Discussion

### 3.1 Perovskite solar cell assembly and encapsulation

Triple mesoscopic PSCs underwent different durations of HTE treatment (Table S1) after conventional annealing at 50ºC for 1.5 hours. Initially, the PSCs exhibited JV curves with the undesired high current shoulder near voltages close to $V_{MPP}$ and larger than the $J_{SC}$. In our hands, the optimal HTE treatment duration was ~40 hours, but all durations (from ~40 to ~120 hours) result in cells with power conversion efficiencies (PCE) within ~7.5 % range. The HTE step does not eliminate the hysteresis observed in all studied devices, where the forward scan (from $J_{SC}$ to $V_{OC}$, FWD JV) is lower in efficiency than scanning from $V_{OC}$ to $J_{SC}$ (BWD JV) as expected in hysteresis-normal devices.[22,47] Epoxy resin encapsulation and clamping cables decreased to ~5% the final PCE due to a deterioration in FF, specifically attributed to larger $R_{SERIES}$ and photocurrent reduction as a result of the encapsulation process. In the Supplementary Information (SI) file, we have included JV curves and box plots showcasing the main PV parameters extracted from the JV curve for all devices after HTE treatment and after tight epoxy encapsulation of the devices (Figure S1-S4).

While these devices might not showcase exceptional efficiency within this particular typology, they demonstrate high reproducibility with commendable and consistent PCEs. The assembly process encompassing pre-laminated glass lids, sealing gaskets, soldered copper cables, attachment of NTC thermistor and embedding in resin epoxy, guarantees the robustness, reproducibility, and stability of these PSC devices, which is a crucial aspect for the MPPT tests outlined in this study.

### 3.2 Validation of the galvanostatic MPPT testing device.

To verify the functionality of our galvanostatic-based MPPT testing device, we selected a commercially available single-crystalline silicon cell (1.18 cm$^2$) and the masked PSCs (0.64 cm$^2$). Figure 2 compares JV curves obtained using the Ossila potentiostat and those acquired with the galvanostatic-based MPPT hardware device developed in this work.

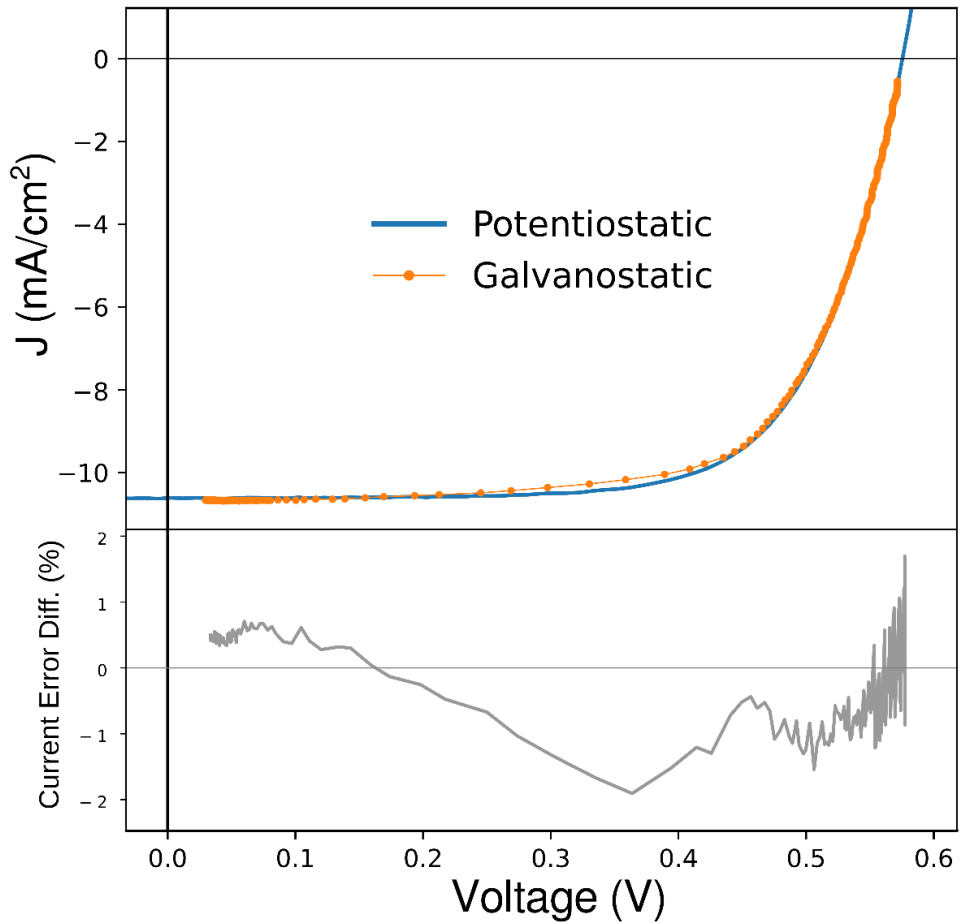

**Figure 2**. Upper panel: JV curve comparison for the Si solar cell recorded using two different techniques: Ossila potentiostat (blue line) and galvanostatic device (orange dotted line). Bottom panel: Percentage error analysis between potentiostatic and galvanostatic methods, with $J_{sc}$ normalized as 100% current reference. Illumination: LED Solar Simulator, 30 mW/cm$^2$.

Galvanostatic and potentiostatic JV curves for the Si-cell device matched well validating our hardware for JV measurements (Support information file contains JV curves and error analysis for 10, 30, 50, 70 and 100 mW/cm$^2$ irradiances, Figure S5).

It is relevant to emphasize that our galvanostatic JV measurement system, while highly effective for certain scenarios explained below, it cannot replace the conventional potentiostatic JV measurement. For instance, the device lacks the capability to measure JV curves in dark conditions. This is because the absence of illumination means that no photocurrent is generated for the photovoltaic device, thus the N-MOSFET acting as variable resistor has no photocurrent to control. Also, $V_{OC}$ and $J_{SC}$ values in these galvanostatic JV curves are extrapolated values because the JV profile never can surpass the JV second quadrant frontier values. Notwithstanding these drawbacks, this *passive* MPP tracker device will excel in measuring light-illuminated JV-SOP cycles for long-term operational stability tests.

In this context of MPPT testing, our focus shifted to the implementation of a basic P&O algorithm including determination of the JV curve of the device under illumination (Figure 3.a), allowing to find the maximum power attainable (Figure 3.b) and its corresponding $V_{MPP}$. The $V_{MPP}$ was then mapped univocally to a specific $V_{GATE}$ to be applied in the N-MOSFET (Figure 3.c). Subsequently, the solar cell device was subjected to the SOP stage setting the before obtained $V_{MPP}$ for 6 min of duration. Figure 3.d.e.f below shows six cycles alternating JV scan and SOP stages in the silicon solar cell connected to our hardware. These cycles were used to record instantaneous current density (Figure 3.d), instantaneous power output (Figure 3.e), and instantaneous cell voltage (Figure 3.f). Throughout this process, the solar cell maintained thermal equilibrium under illumination from a white LED, a state ensured by the attached thermistor (Figure 3.g). This advantageous application of thermistors for temperature sensing would bear particular significance during outdoor experiments, as well as for maintaining precise thermal equilibrium during indoor measurements. Remarkably, our device indirectly sets $V_{MPP}$ in the solar cell by leveraging the microcontroller's stored relationship between the measured voltage in the solar cell (INA219 sensor) and the $V_{GATE}$ applied by the MCP4725 DAC in the gate of the N-MOSFET. But there are other equivalent useful transfer functions relating $V_{GATE}$ with current density, transimpedance or power output from the cell (Figure S6). Overall, this validation test showcases performance and adaptability of our galvanostatic MPPT device for well-behaved solar cells (without hysteresis) offering a promising low-cost MPP tracker or power optimizer for enhancing solar energy harvesting.

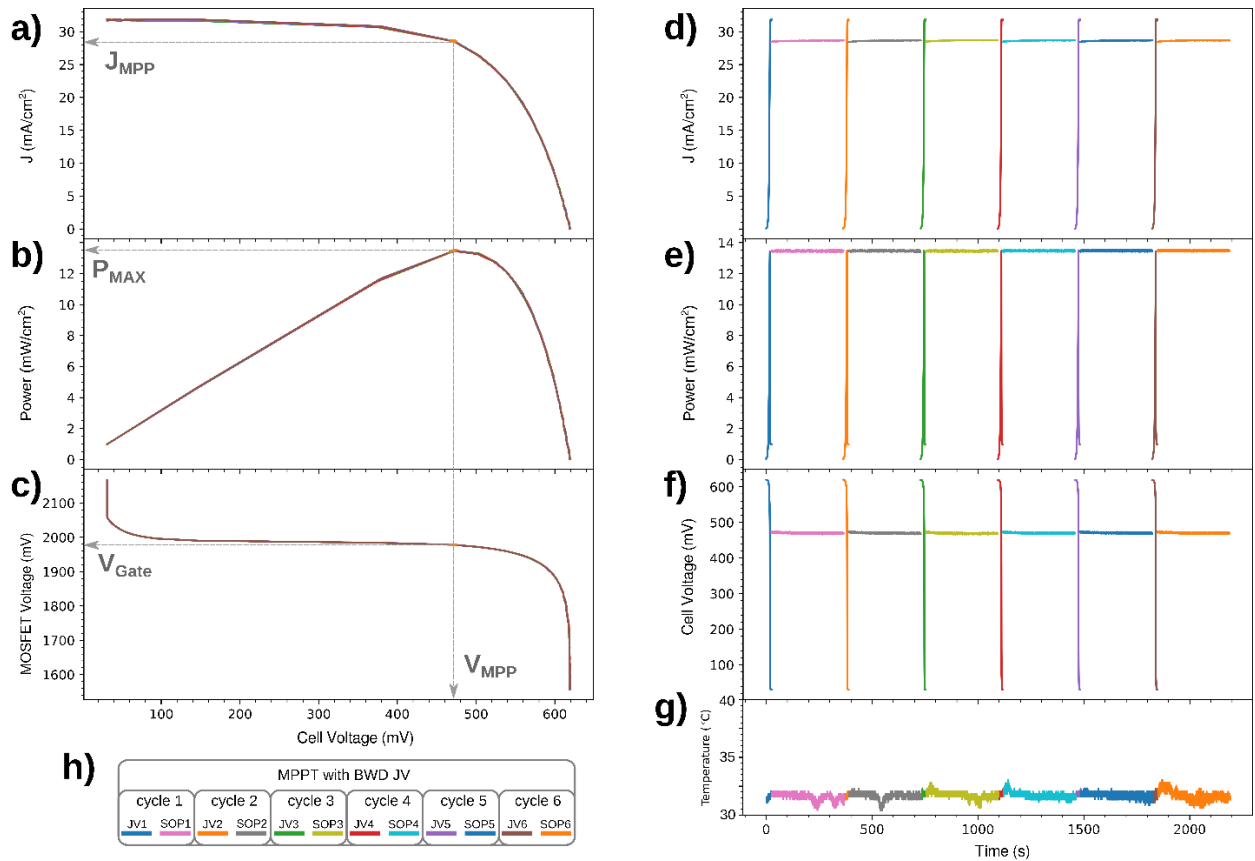

**Figure 3.** a) JV characteristics of the silicon solar cell under constant white LED illumination (~1 Sun), b) power output from the solar cell, c) transfer function between the voltage measured in the solar cell and the voltage applied to the N-MOSFET gate (other transfer functions in Figure S6). Instantaneous d) current density, e) power and f) solar cell voltage during the MPPT sequence alternating J-V curves and SOP stages at $V_{MPP}$, g) solar cell temperature during the MPPT sequence, h) MPPT sequence legend and color code for each stage.

### 3.3 Testing the low-cost galvanostatic MPPT tracker device on PSCs.

After verifying the functionality of our hardware for performing JVs and the basic MPPT algorithm alternating JV-SOP sequence in a silicon solar cell, we applied it to the high hysteresis triple mesoscopic PSC. In comparison to the JVs conducted on the silicon solar cell (Figure 2), noticeable differences were observed between the potentiostatic and galvanostatic JVs, Figure 4.a.

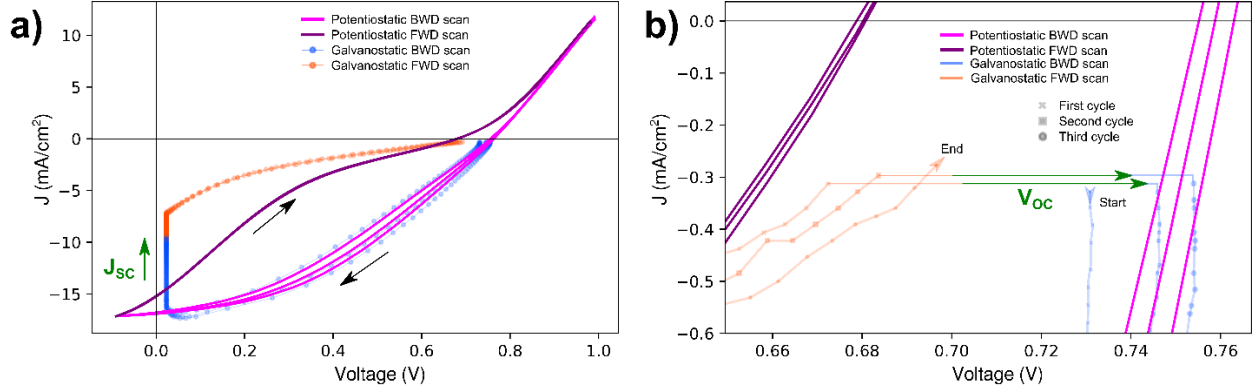

**Figure 4.** a) Three JV cycles from triple mesoscopic or high hysteresis PSCs recorded using a potentiostatic (-) or galvanostatic (-o-) technique. The green arrows highlight the transient dynamics of the recorded $J_{SC}$ during galvanostatic measurement, b) inset for the $V_{OC}$ region in Figure 4.a illustrating the transient dynamics of the $V_{OC}$ during galvanostatic measurement cycles (green arrows).

As observed, both types of measurements capture the hysteresis of the PSC device. The magnitude of hysteresis in PSCs is influenced by scanning speed, settling time, and recent bias history applied in the device.[22,24] The backward (BWD, blue) JV branches of both measurements overlap well; however, the forward (FWD, orange) curve produces a much lower current in the galvanostatic type measurement. The explanation of this effect relies on the effective scan rate and preconditioning time experienced by the PSC during galvanostatic type JV scan compared to potentiostatic one. Conventionally, potentiostatic JV curves can start from the first or fourth quadrant at constant voltage scan rate, inducing the so-called preconditioning stage. However, as explained earlier, our *passive* device cannot impose negative bias value neither voltage higher than $V_{OC}$ in the illuminated solar cell terminals. The open state (low resistance, active state) of the N-MOSFET approximates the solar cell to output a quasi-$J_{SC}$ current density while close state (high resistance, cut-off region) induces quasi-$V_{OC}$ voltage in the cell. The N-MOSFET acting as an electronically controlled variable load records the JV scan during the open-close transient or ohmic region. Indeed, the constant $V_{GATE}$ scan speed implemented in the N-MOSFET does not lead to constant voltage scan rate applied at the cell terminals (Figure S6.c). On other words, it is because the hysteresis index is not a valid metric for quantifying hysteresis in PSCs due to high dependence on JV scan conditions.[48] In the framework of this study and contrarily to well-behaved Si solar cells, the most important aspect is that $J_{SC}$ and $V_{OC}$ are dynamic parameters and it holds paramount significance for the application of broadly implemented fractional $J_{SC}$ and $V_{OC}$ MPPT methods. Figure 4.a clearly shows the decrease of $J_{SC}$ (green arrow) when the N-MOSFET is set to low resistance state. Photocurrent transients and their response shape during on/off light switching are related to charge generation, collection, polarization and recently this response has been ascribed to a delayed photocurrent mode produced by a photoinduced chemical inductor.[49] Recently, a systematic review of the Perovskite Database has uncovered a consistent 4% discrepancy in $J_{SC}$ values derived from JV and EQE measurements.[50] The $J_{SC}$ dynamic behavior during EQE measurements, where device is kept continuously short-circuited, results in a lower integrated $J_{SC}$ value. On the other hand, there is an increase of $V_{OC}$ when the N-MOSFET is set to high resistance or close state, Figure 4.b.

Given the strong significance for the upcoming sections, it is important to note that the presence of substantial hysteresis in the JV creates a loop inside the quadrant. It means that the device can produce identical current levels at two different voltage states, depending on whether it follows the BWD or FWD JV branch. In contrast with the hysteresis observed during the potentiostatic scan, the galvanostatic hysteresis can be said to exhibit a more relaxed behavior. This aspect becomes particularly crucial when attempting to track the MPP of the solar cell using a variable resistor regulating the current output of the device.

Research articles on PSC topic often present JV curves of PSCs cells including the PCE data for both FWD and BWD JV curves as benchmark. However, a crucial challenge arises when attempting to determine the MPP of the cell – should it be based on the BWD or FWD JV curve? In this study, we investigate this ambiguity by employing the conventional potentiostatic approach for JV-SOP cycles of PSC (Figure 5) which later will help to understand the advantages on applying a galvanostatic approach for this task.

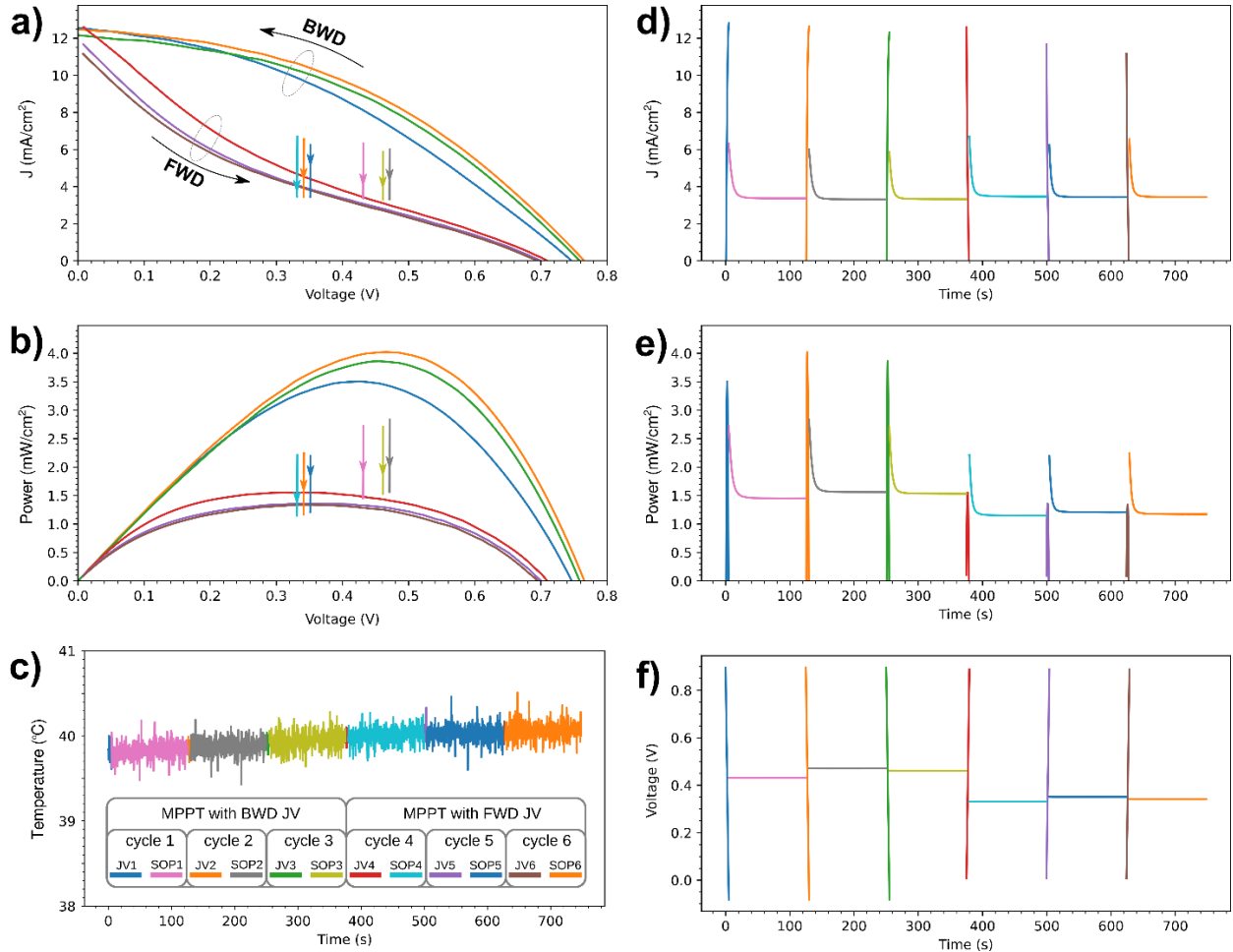

**Figure 5.** Six cycles of MPPT using the potentiostatic approach. First three cycles using a BWD JV for $V_{MPP}$ determination and the last three cycles using FWD JV scans. The SOP stage has 120 seconds of duration. a) JV characteristics of the triple mesoscopic PSC under LED solar simulator (LSH-7320, Newport, AM1.5G, 100 mW/cm²) illumination. SOP stages in this panel appears as vertical lines with the arrowhead pointing the current drifting in time. b) Power output curve from the solar cell during JV scans. SOP stages in this panel appears as vertical lines with the arrowhead point the power output drifting with time. c) Solar cell temperature during the MPPT sequence. This panel contains the legend and color code for all figure with the sequence of cycles, d) instantaneous current during the MPPT sequence alternating J-V curves and SOP stages at $V_{MPP}$, e) instantaneous power during the MPPT sequence alternating JV curves and SOP stages at $V_{MPP}$, f) instantaneous voltage applied in the cell during the MPPT sequence alternating JV curves and SOP stages.

As before with the Si solar cell, the sequence commences by identifying the $V_{MPP}$ after converting the JV curve (Figure 5.a) into a power-voltage curve (Figure 5.b) effectively displaying the peak power achievable. Subsequently, the $V_{MPP}$ value is established within the potentiostat to initiate a SOP stage, which persists for a duration of 120 seconds. Figure 5 shows six cycles, each alternating between JV curve (three BWD + three FWD type) and SOP stages. These cycles were used to record instantaneous current density (Figure 5.d), instantaneous power (Figure 5.e), and voltage applied in the cell by the potentiostat (Figure 5.f). Throughout the process, the device remained in thermal equilibrium (Figure 5.c).

Depending on whether a BWD or FWD type JV scan is conducted, $V_{MPP}$ is ~0.35 V or ~0.45 V, respectively. As expected, the potentiostat fixes the $V_{MPP}$ in the solar cell during the SOP stage. Noteworthy, the highest power peak achieved during the BWD JV stage is not attained during the SOP stage. It progressively declines over time (Figure 5.e and Figure 5.b as indicated with arrows). The decline in power observed during the SOP stage at fixed $V_{MPP}$ is solely attributed to the reduction of the current density output from the cell (Figure 5.d). Interestingly, the current steadily decays until it reaches values comparable to those in the FWD branch during JV tracing stage (Figure 5.a) and power decays until it reaches values comparable to those in the FWD branch of the power-voltage curve (Figure 5.b). Therefore, FWD JV curves offer a more realistic representation of the actual PCE for devices exhibiting significant hysteresis. However, it should be noted that setting the BWD $V_{MPP}$ still results in a power output that is marginally higher compared to $V_{MPP}$ from the FWD branch. This raises two important questions: Is there a possibility of achieving a better PCE from the FWD branch in devices with significant hysteresis? If so, what algorithm could be employed to determine the true MPP of high hysteresis devices?

Instead of the conventional potentiostatic method above, we attach our galvanostatic-based MPPT tracker to drive the PSC utilizing the JV-SOP sequence under illumination, Figure 6. Six galvanostatic JV-SOP cycles alternating $V_{MPP}$ extraction from FWD or BWD type JV curves followed by a SOP stage at $V_{MPP}$. As explained before for JV scan, the galvanostatic method cannot impose a voltage in the solar cell terminals during the SOP stage but control the current flowing out from the cell through the drain/source terminals of the N-MOSFET acting as variable resistor. This variable resistor imposes in the device the load producing the $J_{MPP}$ recorded during the JV stage which at the end fixes the maximum current flowing in the circuit.

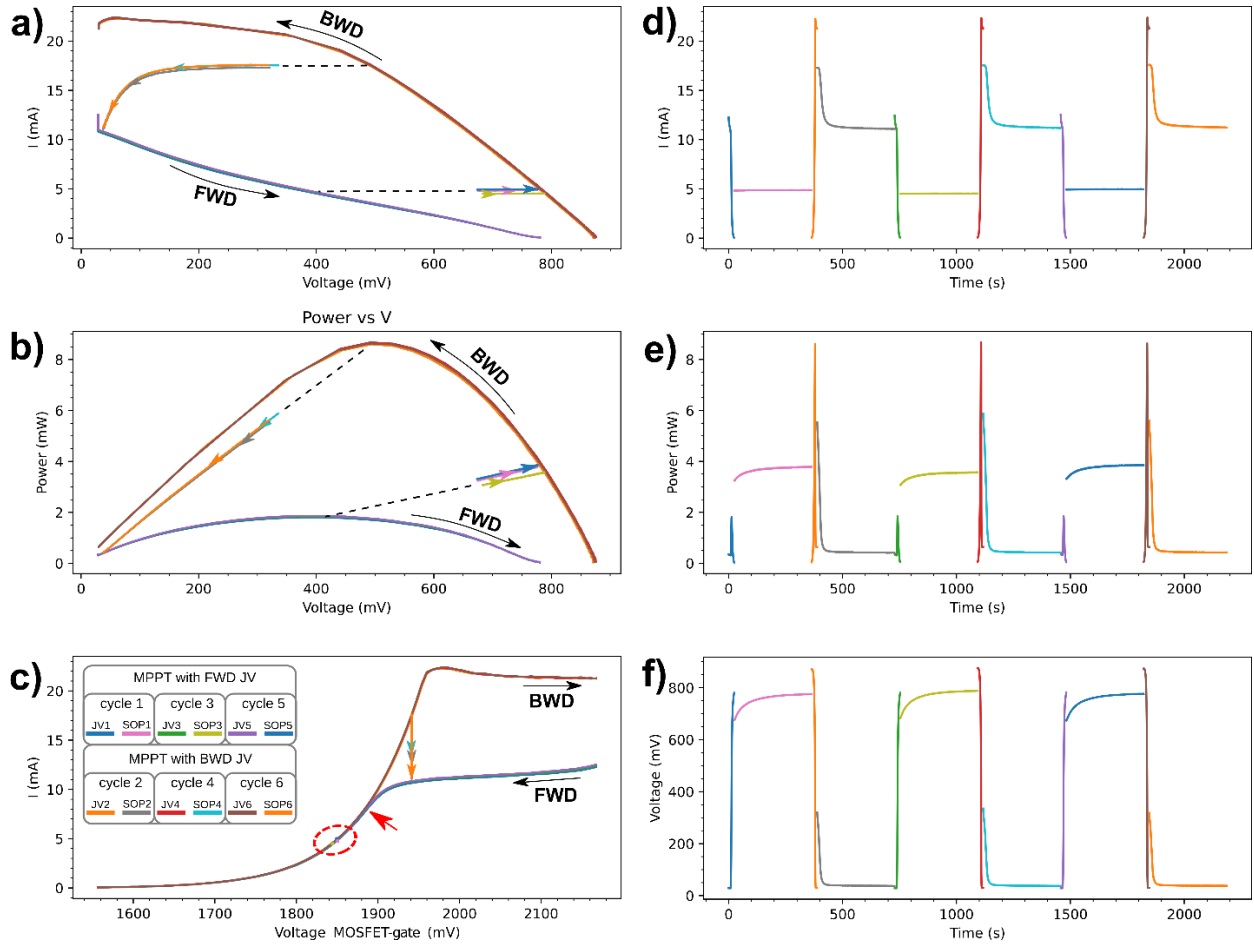

**Figure 6.** Six cycles of MPPT using the galvanostatic approach. Cycles 1, 3 and 5 (2, 4 and 6) use a FWD (BWD) JV scan for $V_{MPP}$ determination. The SOP stage has 360 seconds of duration. a) JV characteristics of the triple mesoscopic PSC under white LED illumination. SOP stages in this panel appears as horizontal lines with the arrowhead pointing the voltage/current drift in time. Dash black line points helps to indicate the $V_{MPP}$ origin. b) Power output curve from the solar cell during JV scans. SOP stages in this panel appears as lines with the arrowhead point the power output drifting with time. Dash black line points to the $V_{MPP}$ origin. c) Solar cell current depending on the voltage applied in the MOSFET gate. SOP stages after BWD JV scan appear as dots (inside the red dash circle). Red arrow indicates the optimal gate voltage to maximize power extraction of the cell. See text for details. This panel contains the legend and color code for all figure with the sequence of JV-SOP cycles, d) instantaneous current, e) power and f) voltage measured in the cell during the MPPT sequence alternating J-V curves and SOP stages.

It is observed that only the cycles using FWD scan type JV offers a stable point of electric power generation during its SOP stage (SOP1, SOP3 and SOP5, Figure 6.b.e) while the MPPT cycles using BWD type JV for $V_{MPP}$ determination, current and voltage decreases to almost no electric power generation during its SOP stage (SOP2, SOP4 and SOP6, Figure 6b.e). This drifting behavior is due to the determined $V_{MPP}$ as starting point for the SOP stage. The transfer function between applied $V_{GATE}$ *vs* current output from device (Figure 6.c) helps to understand this current drift from the device as follows: if the $V_{MPP}$ determined from BWD JV requires a $V_{GATE} > 1900$ mV during the SOP stages, the current output of the device will drift until obtain the corresponding current from the FWD JV (cyan, purple and orange arrows in Figure 6.c). Instead, if the $V_{MPP}$ determined from a FWD JV requires a $V_{GATE} < 1900$ mV, the match of the BWD and FWD branch in the transfer function avoids any current drift (dots surrounded by dashed red circle, Figure 6.c). Interestingly, we demonstrate that using a galvanostatic technique controlling the current output of the device simplifies the MPP tracking process. It results in a considerable increase in power extraction due to the "floating" voltage output naturally increasing from the cell using this technique compared with the conventional potentiostatic MPPT algorithm fixing $V_{MPP}$ during the SOP stage (floating current but it only can decrease as showed in panels Figure 5a.b). As observed in panels Figure 6.b and Figure 6.e, the galvanostatic method produces roughly ~100% increase of the power compared to the obtainable power-voltage FWD curve but is still ~ -50% below that predicted by the power-voltage BWD curve. Note that this improvement relies on the hysteresis grade suffered by the cell. Lower or null hysteresis will imply lower or null improvement.

Due to the evolution to better output powers over time during SOP1, SOP3 and SOP5 stages, the question arose again whether it would be possible to determine an even more optimal $V_{MPP}$ starting point for the SOP stage. Looking in the Figure 6.c, we observe that there is margin to raise the current output of the device by raising the $V_{GATE}$ up to the frontier where BWD and FWD transfer function matches, still avoiding current drift appearing with $V_{GATE}$ > 1900 mV (red arrow, Figure 6.c). First, we try to drive manually the N-MOSFET to reach this point. The Arduino controller outputs the data to the computer using the serial USB connection but also this connection allows inputs "by-hand" to drive the $V_{GATE}$ applied as required in the MOSFET. Then, we can control on-the-fly the variable resistor and consequently the output current from the cell. Figure 7 shows a manual control of this $V_{GATE}$ resulting in an improvement of the power output of the cell. Using this manual control to load the cell, we can drive a SOP stage after FWD JV to increase the power output by ~200% compared to theoretical power from the FWD JV curve (Figure 7.b.e, cycle 1). However, this power output remains below (~ -40%) compared to theoretical power from the BWD JV curve. Interestingly, we can recover manually a failed SOP stage from BWD JV curve to this largest power output produced during cycle 1 (Figure 7.b.e, cycle 2). Figure S7 shows the transfer curves for this manual operation.

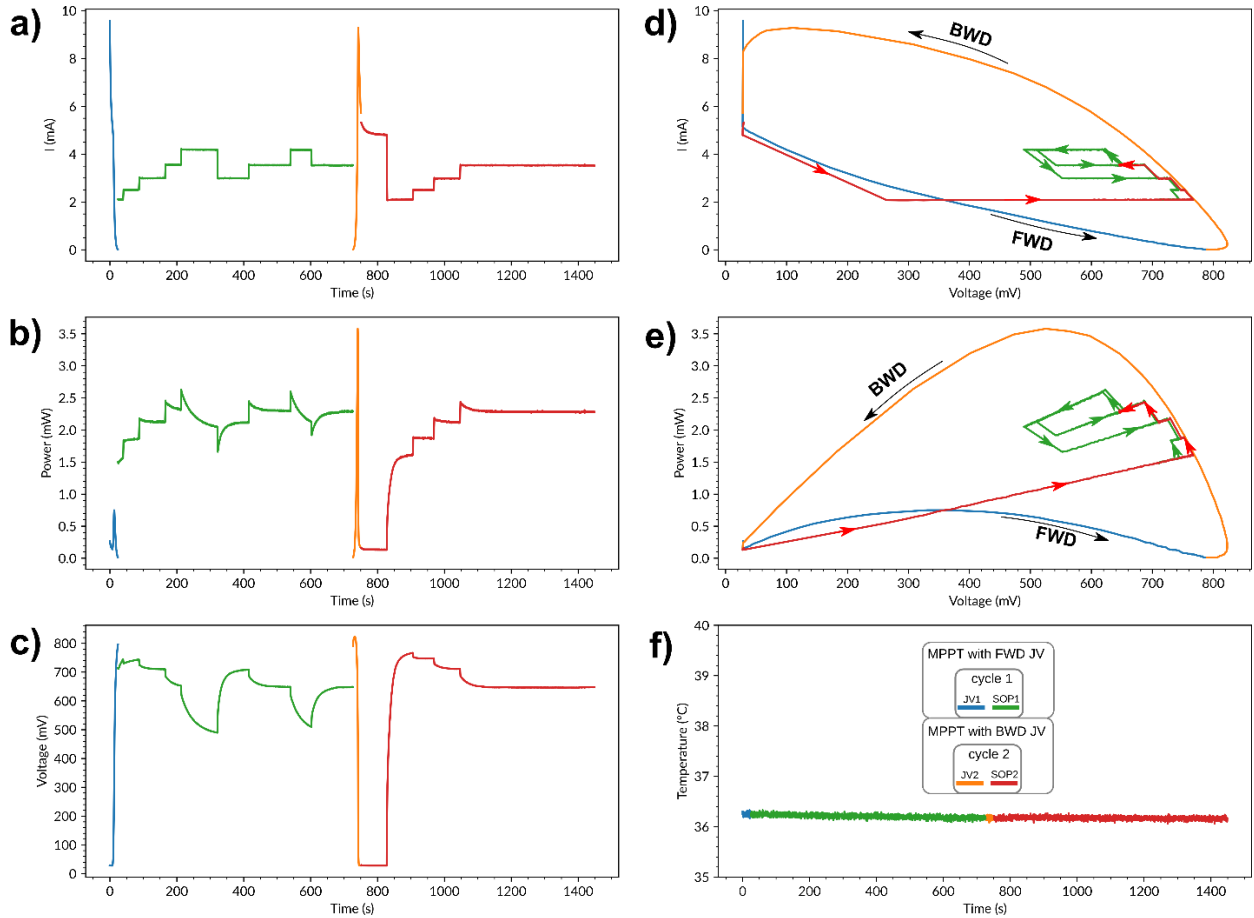

**Figure 7**. Two manual MPPT cycles using the galvanostatic approach. Cycle 1 (2) use a FWD (BWD) JV scan for early $V_{MPP}$ determination. After that, the SOP stages have 720 s of duration and several $V_{GATE}$ input commands by hand are sent to the N-MOSFET to control the output current or active load in the cell terminals. a) Instantaneous current, b) power and c) voltage during the MPPT sequence alternating JV curves and SOP stages under manual control. d) IV curve and e) power output curve from the solar cell during JV scan stage. SOP stages in this panel show a dynamic behavior due the manual control of the $V_{GATE}$ with the arrowhead pointing the voltage/power/current drift time evolution. f) Solar cell temperature during the MPPT sequence. This panel contains the legend and color code for all figure with the sequence of JV-SOP cycles.

The implementation of the algorithm for the automatic search of this optimal $V_{GATE}$ driving the current needs a transfer function that relates $V_{GATE}$ to: i) output current (Figure 6.c), ii) output voltage (Figure 8.a) or iii) power output from cell (Figure 8.b). The MPP search using the transfer function in Figure 6.c is problematic because it requires a threshold describing the frontier point where to discern matched and unmatched FWD and BWD curves. The option of using the transfer function $V_{GATE}$ *vs* output voltage is more promising, Figure 8.a. Here, the optimal $V_{GATE}$ can be found by seeking the voltage at which the maximum difference between the BWD and FWD appears (Figure 8.a, purple line, red arrow). Another procedure to obtain this point is using the transfer function $V_{GATE}$ *vs* cell output power to average the $V_{GATE}$ for each FWD and BWD maximum power points (Figure 8.b, red arrow). This last approach stands out as the most robust, as it remains agnostic regarding the presence or absence of hysteresis within the solar cell. For instance, a Si-cell matches closely both BWD and FWD curves in all their transfer functions due to the absence of hysteresis, Figure S6. Therefore, this optimal point determination is impossible to be found looking for mismatch in the transfer curves. However, the maximum power output is univocally ascribed to the same specific $V_{GATE}$ for both FWD and BWD power curves.

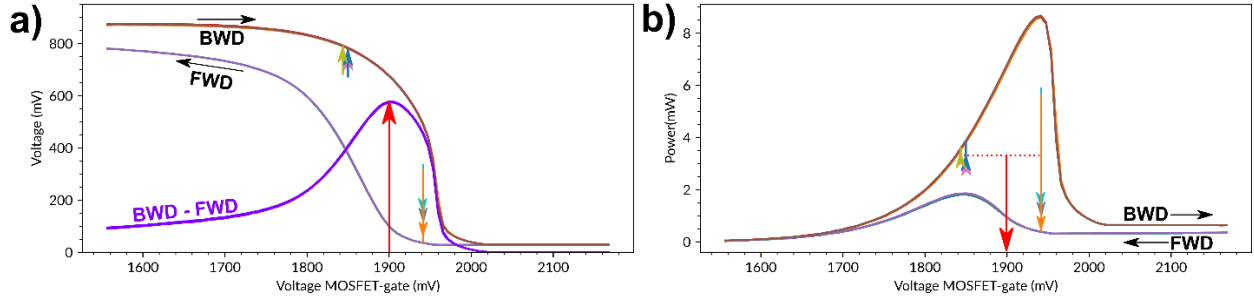

**Figure 8.** a) $V_{GATE}$ *vs* cell voltage and b) $V_{GATE}$ *vs* power output transfer functions. Line color for BWD, FWD JVs and SOP stages follows legend in Figure 6. Red arrow indicates optimal $V_{GATE}$ to drive the output current from the cell. Purple line in Figure 8.a shows the difference between BWD and FWD traces, see details in the text.

Taking all this in mind, the power output *vs* $V_{GATE}$ transfer function serves as the appropriate transfer function for automating the MPP search process as Figure 8b demonstrates. It is noteworthy to mention that Pellet et al.[33] in their work employing a potentiostatic hill-climbing algorithm, discovered that the optimal $V_{MPP}$ slightly surpassed the $V_{MPP}$ derived from JV curves. In contrast, the galvanostatic approach here directly ascertains this slightly above $V_{MPP}$ without the need of iterative procedures. Figure 9 illustrates three JV-SOP cycles implementing the algorithm, each under distinct power illumination conditions, showcasing the automatic search and optimization of $V_{GATE}$ for achieving the MPP.

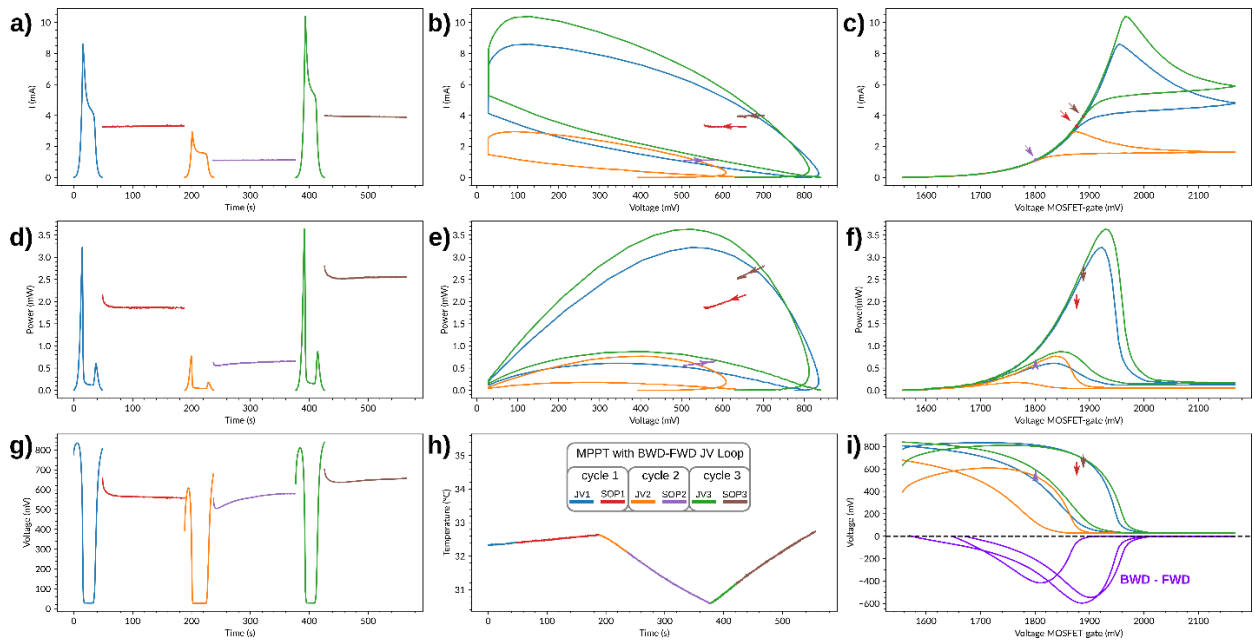

**Figure 9**. Three automatic MPPT search cycles using the galvanostatic approach and power *vs* $V_{GATE}$ transfer curve. Cycles used BWD-FWD JV scan loops for optimal $V_{MPP}$ determination. After that, the SOP stages have 120 s of duration. a) Instantaneous current, b) voltage and c) power during the MPPT sequence alternating JV loop and SOP stages under automatic MPP search. d) JV and e) power-voltage curve output from the solar cell during JV scans. SOP

stages in these panels show minor dynamic behavior due to the automatic control of the $V_{GATE}$ with the arrowhead pointing the voltage/power drift time evolution. f) Solar cell temperature during the MPPT sequence. This panel contains the legend and color code for all figures with the sequence of JV-SOP cycles. $V_{GATE}$ *vs* cell g) current, h) power and i) voltage transfer functions.

Under automatic MPP search of the triple mesoscopic PSC under three different white LED power illumination, the SOP stages still show a lower but overall stable dynamic behavior due to the optimal $V_{GATE}$ found. For the good performance of this automatic search of the MPP in a PSC with high hysteresis, both BWD and FWD JV scan must be determined before SOP stage. Interestingly, we found that it is convenient to scan BWD-FWD JV loops instead of FWD-BWD loops achieves faster the optimal power output because the cell keeps the high voltage state at end of a BWD-FWD JV loop. Figure S8, shows 10 JV-SOP cycles implementing the algorithm with automatic search of the optimal $V_{GATE}$ under the same illumination conditions illustrating the convenience of BWD-FWD JV loops.

### 3.4 Implementation of the algorithm in potentiostatic mode

In principle, there is not a direct way to discover directly and without hill-climbing iteration this optimal output current by using potentiostatic JV measurements. However, we can circumvent this impediment with a workaround making use of the lesson learned using the galvanostatic approach. The proxy method would be to average $J_{MPP,BWD}$ and $J_{MPP,FWD}$ obtained potentiostatically as $J_{MPP,AVG}$ being $VJ(J_{MPP,AVG}) = V_{MPP}$.[51] A graphical derivation of this approach has been deposited as Figure S9.

### 3.5 Implementation of the algorithm for outdoor performance or variable irradiation.

JV-SOP cycles at MPP are generally limited to research contexts for short term stability tests using constant irradiation where JV derivable PV parameters such as $V_{OC}$, $J_{SC}$, FF, $R_{SERIES}$ and $R_{SHUNT}$ need of specific monitoring over time. Practical MPPT algorithms in real-world applications do not require a full JV curve analysis for MPP determination, as this approach is time-consuming and results in suboptimal power production from the solar cell.

In this study, we tested the capability of the microcontroller to drive a well-behaved Si solar cell under one 30-100 mW/cm$^2$ irradiation cycle (EN 50530 standard[43,44]) by implementing a fixed increment P&O MPPT algorithm.[25-28] Our MPPT hardware implementation regulated the cell current output through $V_{GATE}$ modulation, deviating from the traditional practice of controlling the cell voltage output. Despite this difference, this device can run any direct type MPPT algorithms uploaded within the microcontroller. Our test achieved a MPP efficiency ($\eta^{MPP}$) of 98.81%, consistent with typical outcomes of the P&O MPPT algorithm in Si-cells (Figure S10).

We now focus on high hysteresis cells like triple mesoscopic PSCs and use the conventional fixed increment P&O MPPT algorithm guided by the galvanostatic approach. We applied the same 30-100 mW/cm$^2$ irradiation cycle following the EN 50530 standard as before in Si-cell. The results indicates unstable operation and power output oscillation as noted before for PSCs (Figure S11).[32-35] Subsequently, we adapted the standard algorithm to include a feedback validation process ensuring the algorithm's precision in device control. In summary, the typical P&O approach navigates through its decision tree by assessing power output increments at steps *k* and *k+1*. Our adaptation involves reevaluating again step *k* prior to the subsequent cycle to confirm the sign of the $V_{GATE}$ yields an improved power output from the cell (Figure 10).

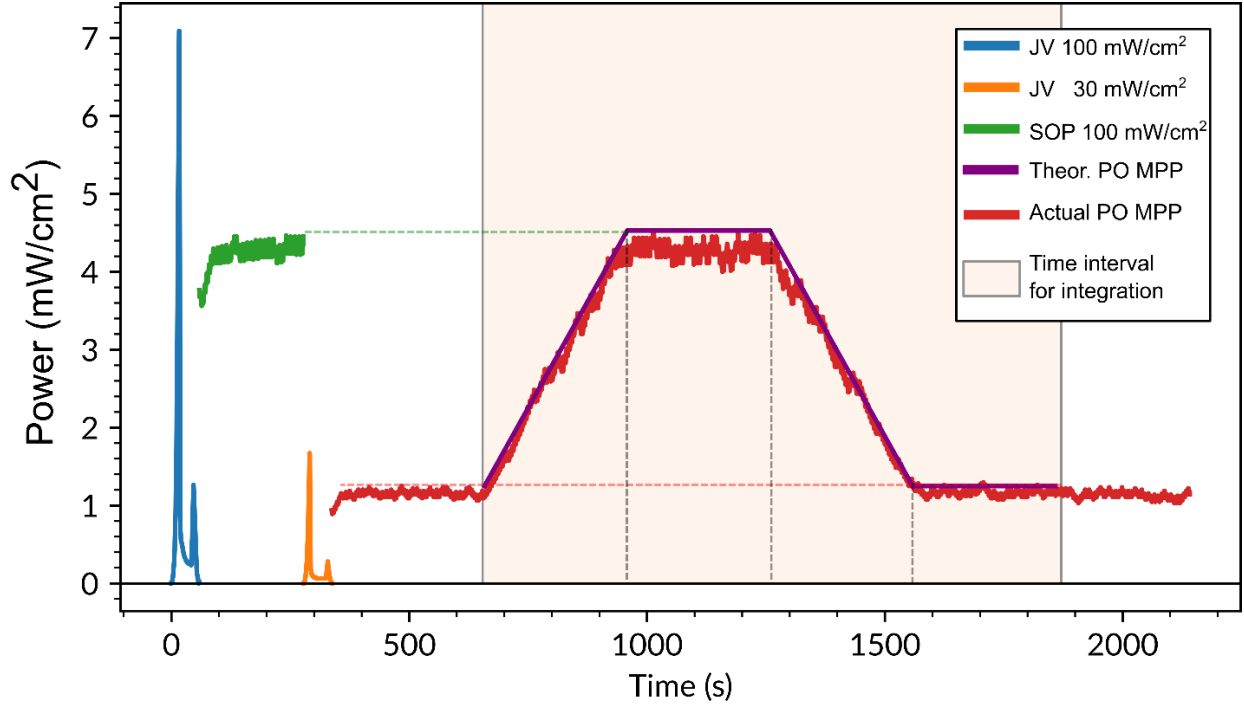

**Figure 10**. Fixed increment P&O MPPT with *k* feedback using the galvanostatic approach for the PSC under one cycle of variable illumination 30/100 mW/cm$^2$ using 300 seconds of ramping and dwelling stages. In blue (orange) the 100 (30) mW/cm$^2$ BWD-FWD JV scan loop for optimal initial $V_{MPP}$ determination. In red (purple) the actual (perfect) instantaneous power output. $\eta^{MPP}$ = 95.02%. $V_{GATE}$ transfer functions in Figure S12.

Based on our understanding, Figure 10 shows for the first time, a P&O MPPT algorithm driving a real PSC (no PV model circuitry) under EN-50530 type variable illumination conditions.

The fast or slow response of our P&O algorithm implemented in our tracker depends on three parameters: 1) number of averaged measurements per data point (β), 2) number of averaged data points constituting a *k*-step (δ), and 3) number of discrete levels of the 12-bits DAC separating two consecutive *k*-steps (ε). One discrete level in our implementation corresponds to 5000/4096 ~ 1.22 mV maximum resolution in the $V_{GATE}$. A variation of approximately ±1 mV in $V_{GATE}$ can lead to a maximum voltage cell change of around ±40 mV in the vicinity of $V_{MPP}$. Conversely, outside the ±40 mV $V_{GATE}$ window centered on $V_{MPP}$, the voltage cell change is limited to ±1 mV or lower. We used β = 60, δ = 10 and ε = 2 for the Si-cell and PSC EN-50530 variable illumination tests above (2.33 W/m$^2$·s slope of variable illumination). The parameters β and δ jointly determine the time required to achieve a *k*-step (overall sampling frequency) while ε influences both the amplitude of inherent oscillations in P&O algorithms and the algorithm's rate of convergence towards the optimal output. A sampling frequency below 200 ms is typically optimal for well-behaved Si and thin-film cells under rapid changes in irradiation. In our algorithm, the decision to increase or decrease $V_{GATE}$ (ε) to track the MPP is made approximately every 2 seconds (β and δ).

Enhancements to this algorithm include the ability to capture rapid changes in illumination through the fine-tuning of β and δ parameters, as well as the exploration of variable ε instead of the current fixed discrete level step.[52] It is important to note that these advancements extend beyond the current scope of this study. Our ongoing research is focused on this optimization of the algorithm and deploying this tracking device to study long-term operational stability of PSCs under outdoor conditions.

## 4. Conclusions

In this study, we developed a galvanostatic MPPT device aimed at enhancing the PCE of high hysteresis PSCs more rapidly compared to the potentiostatic approach. The algorithm and hardware introduced in this work have the potential to democratize and streamline both outdoor and short-term operational stability measurements in the laboratory, propelling the advancement of emerging photovoltaics characterized by substantial J-V hysteresis. Moreover, the integration of galvanostatic power optimizers employing this MPP search algorithm with the solar panel prior to the inverter could potentially enable the commercial viability of extremely low-cost carbon-based PSCs afflicted by high hysteresis. This device allows cost-effective testing of solar cells, with an estimated cost of approximately 10 USD per solar cell. The primary cost contributors include the microcontroller, voltage-current sensor, and DAC. The resulting device is compact, portable, and easily integrable into a glove box or for unattended outdoor measurements, featuring provisions for a memory card for data storage and a battery. The circuitry incorporates device temperature (or ambient temperature) data recording using a thermistor. On the firmware side, we utilized an open-source electronics platform, specifically Arduino, to facilitate user-friendly hardware and software modifications. Additionally, leveraging an open-source language like Python, along with associated libraries for data collection, storage, calculation, and graphing, promotes extensive user engagement, encouraging continuous development and enhancement opportunities.

## 5. Acknowledgments

E.J.J.-P. acknowledges the funding support from MCIN/AEI/ 10.13039/501100011033 and European Union NextGenerationEU/ PRTR (project grants PID2019-107893RB-I00 and EIN2020-112315, respectively). C. M. thanks the grant IJC2020-044684-I funded by MCIN/AEI/10.13039/501100011033 and by European Union NextGenerationEU/PRTR. M.H. acknowledges the funding support from MCIN/AEI/10.13039/501100011033 for the Ramón y Cajal fellowship (RYC-2018-025222-I) and the project PID2019-108247RA-I00. M.H also acknowledges the DGA/fondos FEDER (construyendo Europa desde Aragón) for funding the research group Platon (E31_20R) and the project LMP71_21.

## 6. Declaration of Interests

The authors declare no competing interests.

# 7. References

(1) Best Research-Cell Efficiency Chart from NREL (Sept. 2023). https://www.nrel.gov/pv/assets/pdfs/best-research-cell-efficiencies.pdf

(2) Bisquert, J.; Juarez-Perez, E. J. The Causes of Degradation of Perovskite Solar Cells. *J. Phys. Chem. Lett.* **2019**, *10*, 5889-5891.

(3) Juarez-Perez, E. J.; Ono, L. K.; Qi, Y. Thermal degradation of formamidinium based lead halide perovskites into sym-triazine and hydrogen cyanide observed by coupled thermogravimetry - mass spectrometry analysis. *J. Mater. Chem. A* **2019**, *7*, 16912-16919.

(4) Juarez-Perez, E. J.; Ono, L. K.; Uriarte, I.; Cocinero, E. J.; Qi, Y. Degradation Mechanism and Relative Stability of Methylammonium Halide Based Perovskites Analyzed on the Basis of Acid-Base Theory. *ACS Appl. Mater. Interfaces* **2019**, *11*, 12586-12593.

(5) García-Fernández, A.; Juarez-Perez, E. J.; Castro-García, S.; Sánchez-Andújar, M.; Ono, L. K.; Jiang, Y.; Qi, Y. Benchmarking chemical stability of arbitrarily mixed 3D hybrid halide perovskites for solar cell applications. *Small Methods* **2018**, 1800242.

(6) Juarez-Perez, E. J.; Ono, L. K.; Maeda, M.; Jiang, Y.; Hawash, Z.; Qi, Y. B. Photodecomposition and thermal decomposition in methylammonium halide lead perovskites and inferred design principles to increase photovoltaic device stability. *J. Mater. Chem. A* **2018**, *6*, 9604-9612.

(7) Shi, L.; Bucknall, M. P.; Young, T. L.; Zhang, M.; Hu, L.; Bing, J.; Lee, D. S.; Kim, J.; Wu, T.; Takamure, N.; McKenzie, D. R.; Huang, S.; Green, M. A.; Ho-Baillie, A. W. Y. Gas chromatography-mass spectrometry analyses of encapsulated stable perovskite solar cells. *Science* **2020**, *368*, 10.1126/science.aba2412.

(8) Dunfield, S. P.; Louks, A. E.; Waxse, J.; Tirawat, R.; Robbins, S.; Berry, J. J.; Reese, M. O. Forty-two days in the SPA, building a stability parameter analyzer to probe degradation mechanisms in perovskite photovoltaic devices. *Sustainable Energy &mathsemicolon Fuels* **2023**, *7*, 3294-3305.

(9) Jiang, Q.; Tirawat, R.; Kerner, R. A.; Gaulding, E. A.; Xian, Y.; Wang, X.; Newkirk, J. M.; Yan, Y.; Berry, J. J.; Zhu, K. Towards linking lab and field lifetimes of perovskite solar cells. *Nature* **2023**, 10.1038/s41586-41023.

(10) Duan, M.; Hu, Y.; Mei, A.; Rong, Y.; Han, H. Printable carbon-based hole-conductor-free mesoscopic perovskite solar cells: From lab to market. *Materials Today Energy* **2018**, *7*, 221-231.

(11) Domanski, K.; Correa-Baena, J.-P.; Mine, N.; Nazeeruddin, M. K.; Abate, A.; Saliba, M.; Tress, W.; Hagfeldt, A.; Grätzel, M. Not All That Glitters Is Gold: Metal-Migration-Induced Degradation in Perovskite Solar Cells. *ACS Nano* **2016**, *10*, 6306-6314.

(12) Juárez-Pérez, E. J.; Leyden, M. R.; Wang, S.; Ono, L. K.; Hawash, Z.; Qi, Y. Role of the Dopants on the Morphological and Transport Properties of Spiro-MeOTAD Hole Transport Layer. *Chem. Mater.* **2016**, *28*, 5702-5709.

(13) He, S.; Qiu, L.; Son, D.-Y.; Liu, Z.; Juarez-Perez, E. J.; Ono, L. K.; Stecker, C.; Qi, Y. Carbon-based Electrode Engineering Boosts the Efficiency of All Low-temperature Processed Perovskite Solar Cells. *ACS Energy Letters* **2019**, *4*, 2032-2039.

(14) Snaith, H. J.; Abate, A.; Ball, J. M.; Eperon, G. E.; Leijtens, T.; Noel, N. K.; Stranks, S. D.; Wang, J. T.-W.; Wojciechowski, K.; Zhang, W. Anomalous Hysteresis in Perovskite Solar Cells. *Journal of Physical Chemistry Letters* **2014**, *5*, 1511-1515.

(15) Kim, H.-S.; Park, N.-G. Parameters Affecting I – V Hysteresis of CH3NH3PbI3 Perovskite Solar Cells: Effects of Perovskite Crystal Size and Mesoporous TiO2 Layer. *J. Phys. Chem. Lett.* **2014**, *5*, 2927-2934.

(16) Seol, D.; Jeong, A.; Han, M. H.; Seo, S.; Yoo, T. S.; Choi, W. S.; Jung, H. S.; Shin, H.; Kim, Y. Origin of Hysteresis in CH3NH3PbI3Perovskite Thin Films. *Adv. Funct. Mater.* **2017**, *27*, 1701924.

(17) Almosni, S.; Cojocaru, L.; Li, D.; Uchida, S.; Kubo, T.; Segawa, H. Tunneling-Assisted Trapping as one of the Possible Mechanisms for the Origin of Hysteresis in Perovskite Solar Cells. *Energy Technology* **2017**.

(18) Li, C.; Guerrero, A.; Zhong, Y.; Huettner, S. Origins and mechanisms of hysteresis in organometal halide perovskites. *Journal of Physics Condensed Matter* **2017**, *29*.

(19) Song, D. H.; Jang, M. H.; Lee, M. H.; Heo, J. H.; Park, J. K.; Sung, S.-J.; Kim, D.-H.; Hong, K.-H.; Im, S. H. A discussion on the origin and solutions of hysteresis in perovskite hybrid solar cells. *J. Phys. D: Appl. Phys.* **2016**, *49*, 473001.

(20) Wu, Y.; Shen, H.; Walter, D.; Jacobs, D.; Duong, T.; Peng, J.; Jiang, L.; Cheng, Y.-B.; Weber, K. On the Origin of Hysteresis in Perovskite Solar Cells. *Adv. Funct. Mater.* **2016**.

(21) Chen, B.; Yang, M.; Priya, S.; Zhu, K. b. Origin of J-V Hysteresis in Perovskite Solar Cells. *J. Phys. Chem. Lett.* **2016**, *7*, 905-917.

(22) Liu, P.; Wang, W.; Liu, S.; Yang, H.; Shao, Z. Fundamental understanding of photocurrent hysteresis in perovskite solar cells. *Advanced Energy Materials* **2019**, *9*, 1803017.

(23) Cai, F.; Yang, L.; Yan, Y.; Zhang, J.; Qin, F.; Liu, D.; Cheng, Y.-B.; Zhou, Y.; Wang, T. Eliminated hysteresis and stabilized power output over 20% in planar heterojunction perovskite solar cells by compositional and surface modifications to the low-temperature-processed TiO2 layer. *J. Mater. Chem. A* **2017**, *5*, 9402-9411.

(24) Unger, E.; Paramasivam, G.; Abate, A. Perovskite solar cell performance assessment. *Journal of Physics: Energy* **2020**, *2*, 044002.

(25) Selvan, S.; Nair, P.; Umayal, U. A Review on Photo Voltaic MPPT Algorithms. *International Journal of Electrical and Computer Engineering (IJECE)* **2016**, *6*, 567.

(26) Verma, D.; Nema, S.; Shandilya, A. M.; Dash, S. K. Comprehensive analysis of maximum power point tracking techniques in solar photovoltaic systems under uniform insolation and partial shaded condition. *Journal of Renewable and Sustainable Energy* **2015**, *7*, 042701.

(27) Sharma, C.; Jain, A. Maximum Power Point Tracking Techniques: A Review. *International Journal of Recent Research in Electrical and Electronics Engineering* **2014**, *1*, 25-33.

(28) Babaa, S. E.; Armstrong, M.; Pickert, V. Overview of Maximum Power Point Tracking Control Methods for PV Systems. *Journal of Power and Energy Engineering* **2014**, *02*, 59-72.

(29) Czudek, A.; Hirselandt, K.; Kegelmann, L.; Al-Ashouri, A.; Jošt, M.; Zuo, W.; Abate, A.; Korte, L.; Albrecht, S.; Dagar, J.; Unger, E. L. Transient Analysis during Maximum Power Point Tracking (TrAMPPT) to Assess Dynamic Response of Perovskite Solar Cells. *Arxiv* **2019**, 1-23.

(30) Dunbar, R. B.; Duck, B. C.; Moriarty, T.; Anderson, K. F.; Duffy, N. W.; Fell, C. J.; Kim, J.; Ho-Baillie, A.; Vak, D.; Duong, T.; Wu, Y.; Weber, K.; Pascoe, A.; Cheng, Y.-B.; Lin, Q.; Burn, P. L.; Bhattacharjee, R.; Wang, H.; Wilson, G. J. How reliable are efficiency measurements of perovskite solar cells? the first inter-comparison, between two accredited and eight non-accredited laboratories. *J. Mater. Chem. A* **2017**, *5*, 22542-22558.

(31) Song, T.; Ottoson, L.; Gallon, J.; Friedman, D. J.; Kopidakis, N. Reliable Power Rating of Perovskite PV Modules. *2021 IEEE 48th Photovoltaic Specialists Conference (PVSC)* **2021**.

(32) Cimaroli, A. J.; Yu, Y.; Wang, C.; Liao, W.; Guan, L.; Grice, C. R.; Zhao, D.; Yan, Y. Tracking the maximum power point of hysteretic perovskite solar cells using a predictive algorithm. *J. Mater. Chem. C* **2017**, *5*, 10152-10157.

(33) Pellet, N.; Giordano, F.; Ibrahim Dar, M.; Gregori, G.; Zakeeruddin, S. M.; Maier, J.; Grätzel, M. Hill climbing hysteresis of perovskite-based solar cells: A maximum power point tracking investigation. *Progress in Photovoltaics: Research and Applications* **2017**.

(34) Rakocevic, L.; Ernst, F.; Yimga, N. T.; Vashishtha, S.; Aernouts, T.; Heumueller, T.; Brabec, C. J.; Gehlhaar, R.; Poortmans, J. Reliable performance comparison of perovskite solar cells using optimized maximum power point tracking. *Solar Rrl* **2019**, *3*, 1800287.

(35) Saito, H.; Aoki, D.; Tobe, T.; Magaino, S. Development of a New Maximum Power Point Tracking Method for Power Conversion Efficiency Measurement of Metastable Perovskite Solar Cells. *Electrochemistry* **2020**, *88*, 218-223.

(36) Köbler, H.; Neubert, S.; Jankovec, M.; Glažar, B.; Haase, M.; Hilbert, C.; Topič, M.; Rech, B.; Abate, A. High-Throughput Aging System for Parallel Maximum Power Point Tracking of Perovskite Solar Cells. *Energy Technology* **2022**, *10*, 2200234.

(37) Abdelwanis, M. I.; Zaky, A. A.; Ali, M. M. Performance study of linear induction motor fed from perovskite solar cells based on GA MPPT. *2022 23rd International Middle East Power Systems Conference (MEPCON)* **2022**.

(38) Olzhabay, Y.; Ng, A.; Ukaegbu, I. A. Perovskite PV Energy Harvesting System for Uninterrupted IoT Device Applications. *Energies* **2021**, *14*, 7946.

(39) https://www.arduino.cc/

(40) Hashmi, S. G.; Martineau, D.; Dar, M. I.; Myllymäki, T. T. T.; Sarikka, T.; Ulla, V.; Zakeeruddin, S. M.; Grätzel, M. High performance carbon-based printed perovskite solar cells with humidity assisted thermal treatment. *J. Mater. Chem. A* **2017**, *5*, 12060-12067.

(41) Young, J. F. Humidity control in the laboratory using salt solutions—a review. *Journal of Applied Chemistry* **1967**, *17*, 241-245.

(42) Grapa "GRAphing and Photovoltaics Analysis". Romain Carron. 10.5281/zenodo.1164571

(43) Andrejasvicv, T.; Jankovec, M.; Topicv, M. Comparison of direct maximum power point tracking algorithms using EN 50530 dynamic test procedure. *IET Renewable Power Generation* **2011**, *5*, 281.

(44) Bründlinger, R.; Henze, N.; Häberlin, H.; Burger, B.; Bergmann, A.; Baumgartner, F. prEN 50530–The new European standard for performance characterisation of PV inverters. *24th EU PV Conf., Hamburg, Germany* **2009**.

(45) Duran, E.; Piliougine, M.; Sidrach-de-Cardona, M.; Galan, J.; Andujar, J. M. Different methods to obtain the I--V curve of PV modules: A review. *33rd IEEE Photovoltaic Specialists Conference* **2008**, 1-6.

(46) https://github.com/ej-jp/perovskino

(47) Rong, Y.; Hu, Y.; Ravishankar, S.; Liu, H.; Hou, X.; Sheng, Y.; Mei, A.; Wang, Q.; Li, D.; Xu, M.; Bisquert, J.; Han, H. Tunable hysteresis effect for perovskite solar cells. *Energy and Environmental Science* **2017**, *10*, 2383-2391.

(48) Habisreutinger, S. N.; Noel, N. K.; Snaith, H. J. Hysteresis Index: A Figure without Merit for Quantifying Hysteresis in Perovskite Solar Cells. *ACS Energy Lett.* **2018**, *3*, 2472-2476.

(49) Bisquert, J.; Gonzales, C.; Guerrero, A. Transient On/Off Photocurrent Response of Halide Perovskite Photodetectors. *J. Phys. Chem. C* **2023**, 0.

(50) Saliba, M.; Unger, E.; Etgar, L.; Luo, J.; Jacobsson, T. J. A systematic discrepancy between the short circuit current and the integrated quantum efficiency in perovskite solar cells. *Nature Communications* **2023**, *14*, 14:5445.

(51) VJ is the inverse function of JV.

(52) Soetedjo, A.; Sulistiawati, I. B.; Nakhoda, Y. I. A Modified Step Size Perturb and Observe Maximum Power Point Tracking for PV System. *IEEE International Conference on Engineering, Science, and Industrial Applications (ICESI)* **2019**, 1-6.

# Support Information:
# Enhanced Power Point Tracking for High Hysteresis Perovskite Solar Cells: A Galvanostatic Approach

Emilio J. Juarez-Perez[1,2*], Cristina Momblona[1], Roberto Casas[3], Marta Haro[1,4]

1: Nanostructured Films & Particles Research Group (NFP). Instituto de Nanociencia y Materiales de Aragón (INMA). CSIC-Universidad de Zaragoza, Zaragoza 50009, Spain.

2: Aragonese Foundation for Research and Development (ARAID) Government of Aragon. Zaragoza 50018, Spain

3: Howlab - Human Openware Research Group, Aragon Institute of Engineering Research (I3A), University of Zaragoza, 50018 Zaragoza, Spain

4: Departamento de Química Física. Facultad de Ciencias. Universidad de Zaragoza. Zaragoza 50009, Spain

Corresponding Author: ejjuarezperez@unizar.es

## Contents

# 1. Solaronix triple mesoscopic perovskite solar cells characterization.

**Table S1**. Naming convention for the triple mesoscopic perovskite solar cells

| Device Name | HTE duration (h) |
|:---:|:---:|
| Solaronix4 | 110 |
| Solaronix5 | 70 |
| Solaronix6 | 41 |
| Solaronix7 | 118 |

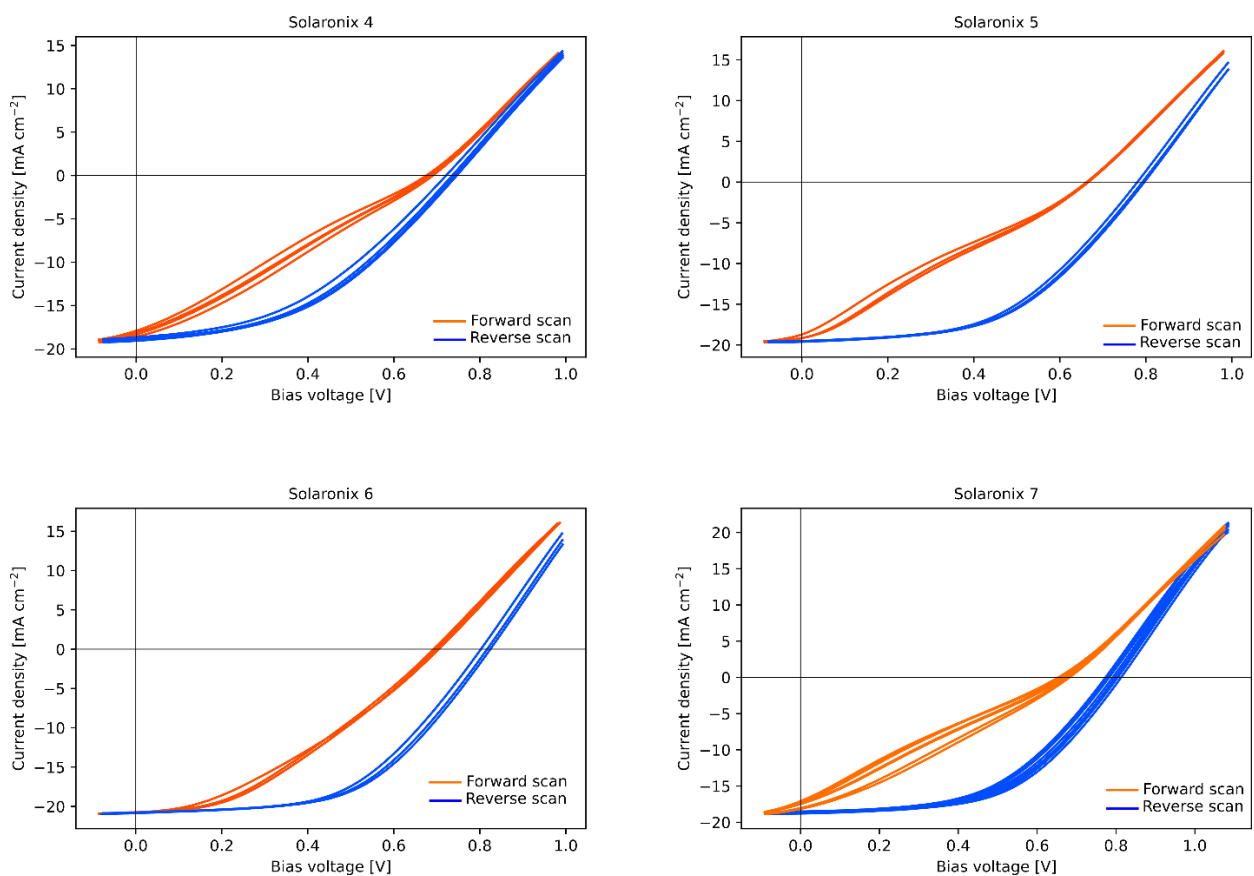

**Figure S1**. JV curves for Solaronix perovskite solar cells after HTE treatment (100 mW/cm$^2$).

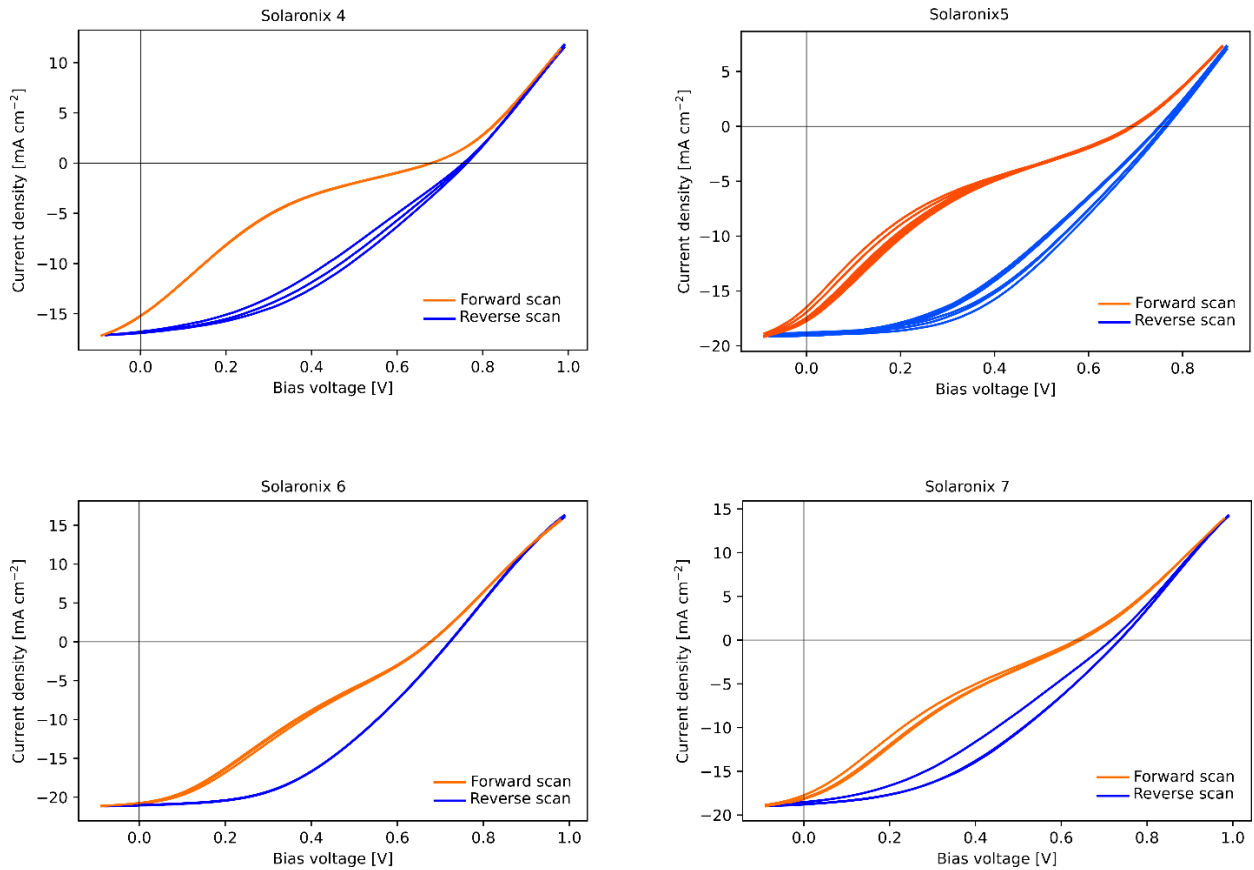

**Figure S2**. JV curves for Solaronix perovskite solar cells after complete encapsulation of the device (100 mW/cm$^2$).

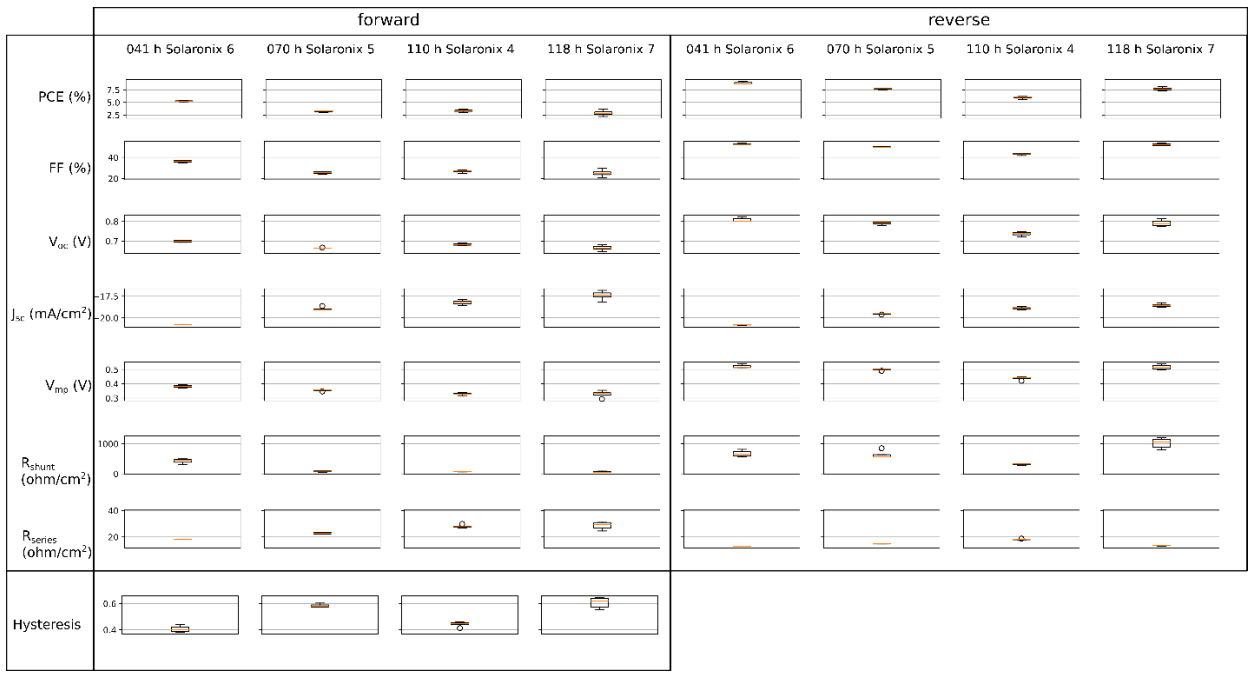

**Figure S3**. Box plots showcasing key PV parameters extracted from the corresponding JV curves for the different Solaronix perovskite solar cells after post HTE treatment (100 mW/cm$^2$).

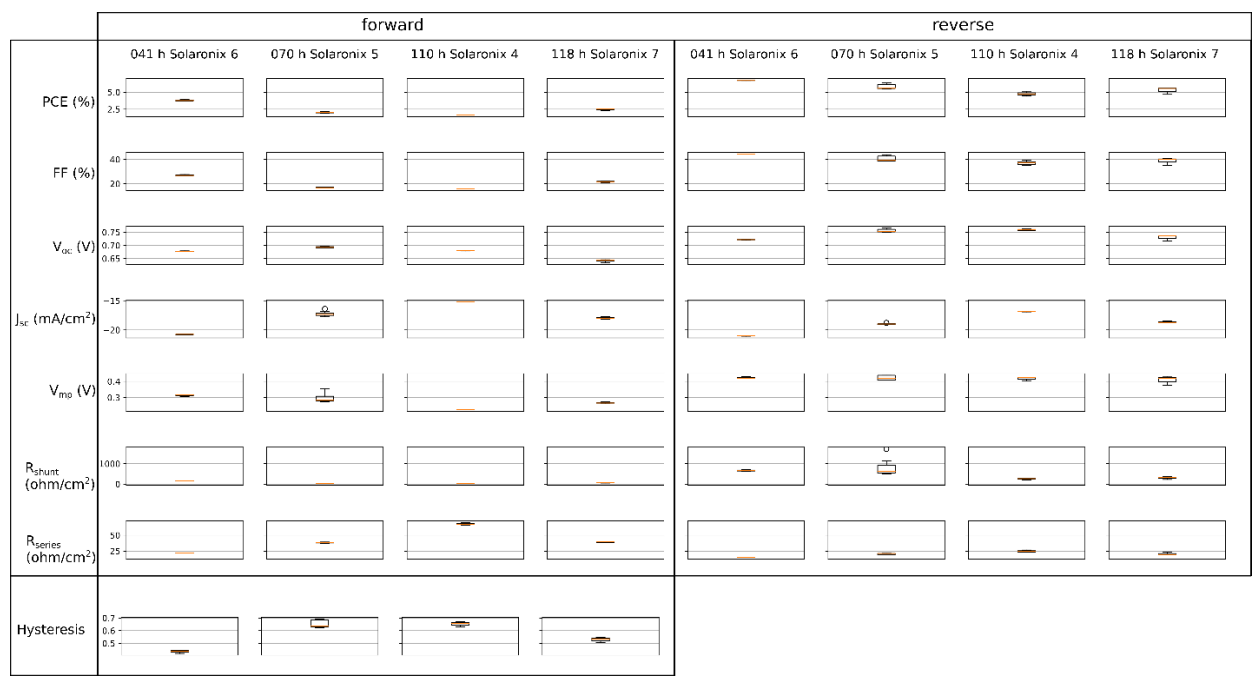

**Figure S4**. Box plots showcasing key PV parameters extracted from the corresponding JV curves for the different Solaronix perovskite solar cells after complete encapsulation of the device (100 mW/cm$^2$).

# 2. Validation of the galvanostatic MPPT testing device running a silicon solar cell.

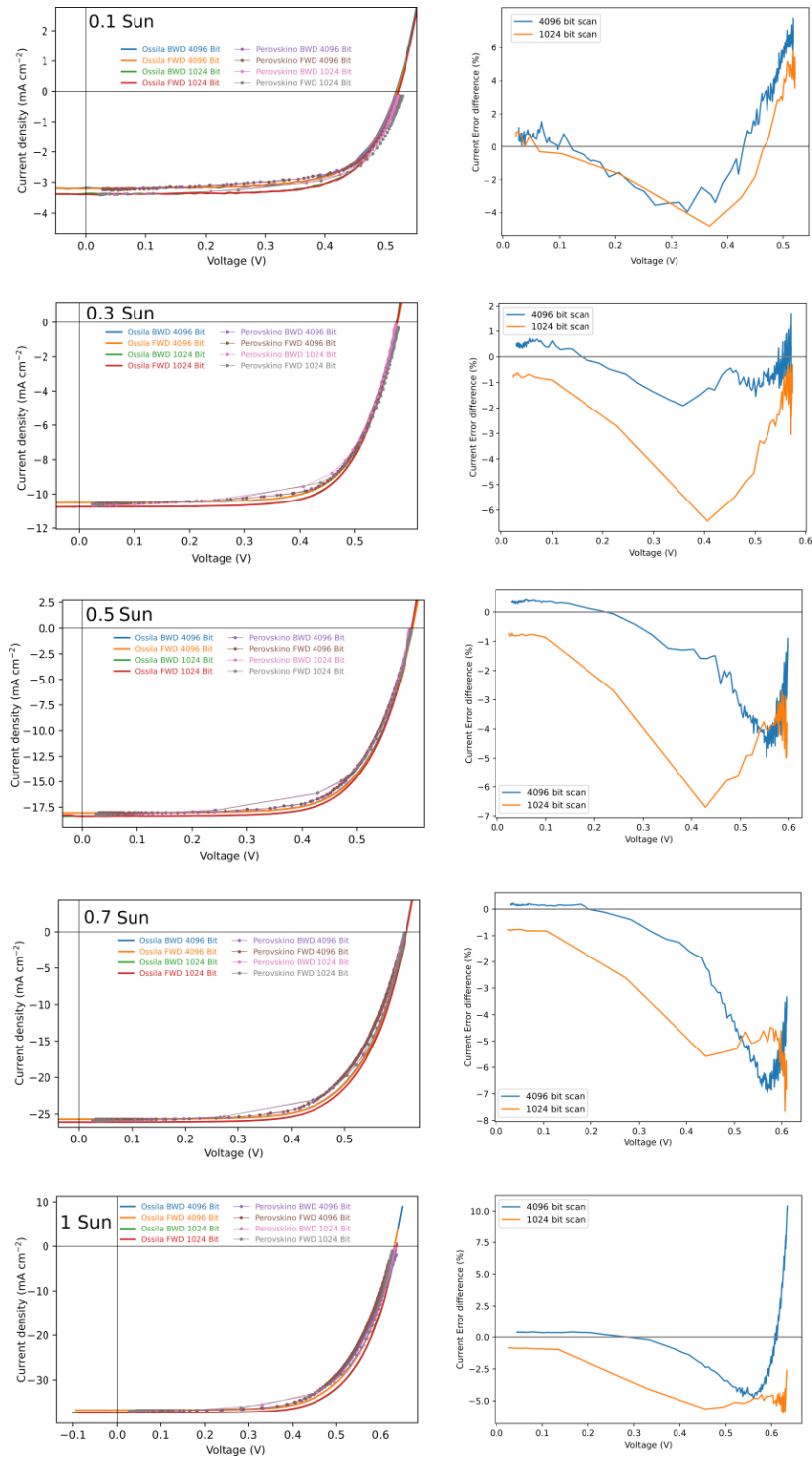

**Figure S5**. Left panels: JV curves comparison for the Si solar cell recorded using two different techniques: Ossila Potentiostat (lines) and galvanostatic device (dotted lines). Right panels: Percentage error analysis between potentiostatic and galvanostatic methods, with $J_{sc}$ normalized as 100% current reference.

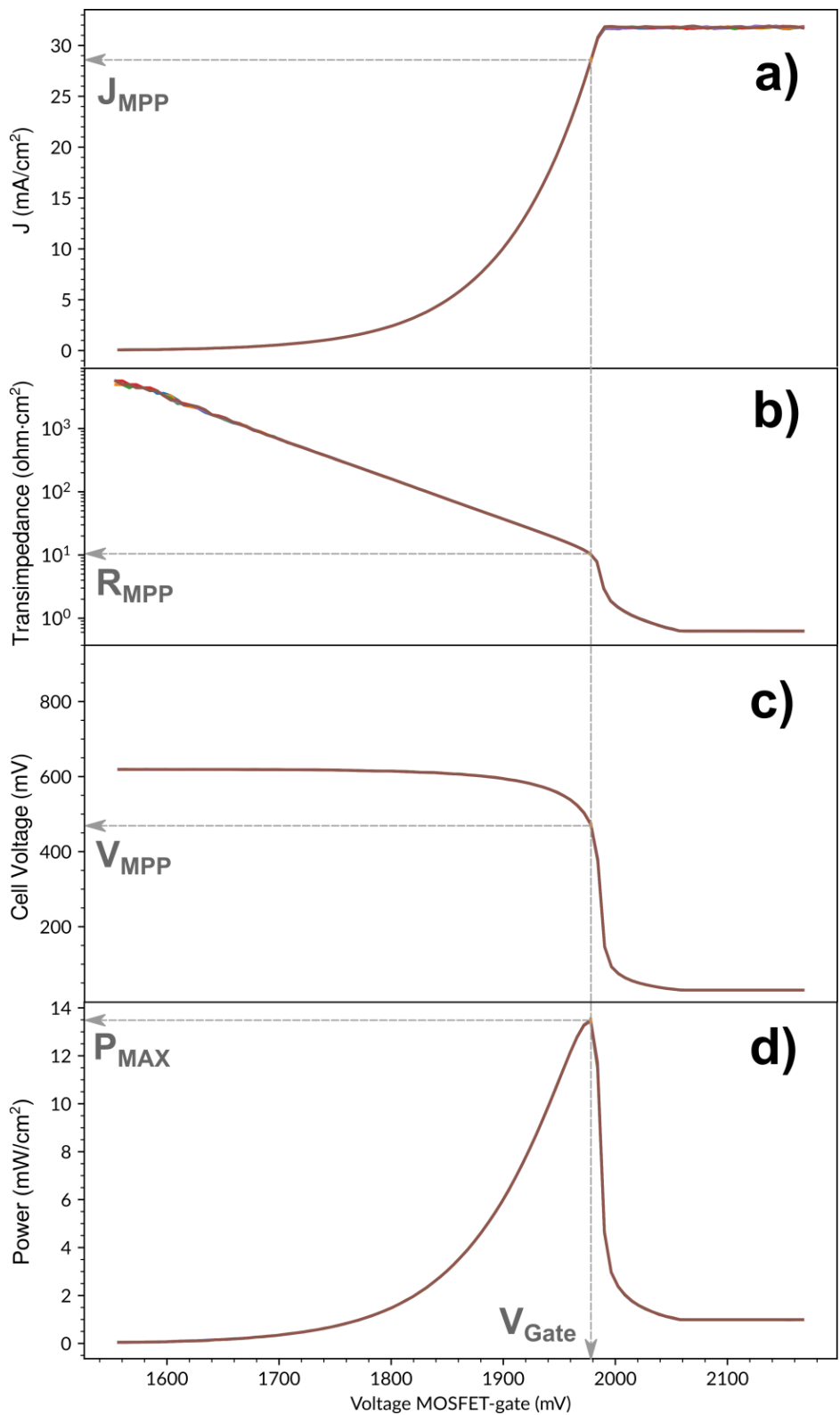

**Figure S6**. V$_{GATE}$ vs a) cell current density, b) transimpedance, c) cell voltage and d) power output transfer functions. Gray arrows indicate optimal V$_{GATE}$ to drive the cell to MPP.

## 3. Testing the galvanostatic MPPT tracker device on PSCs.

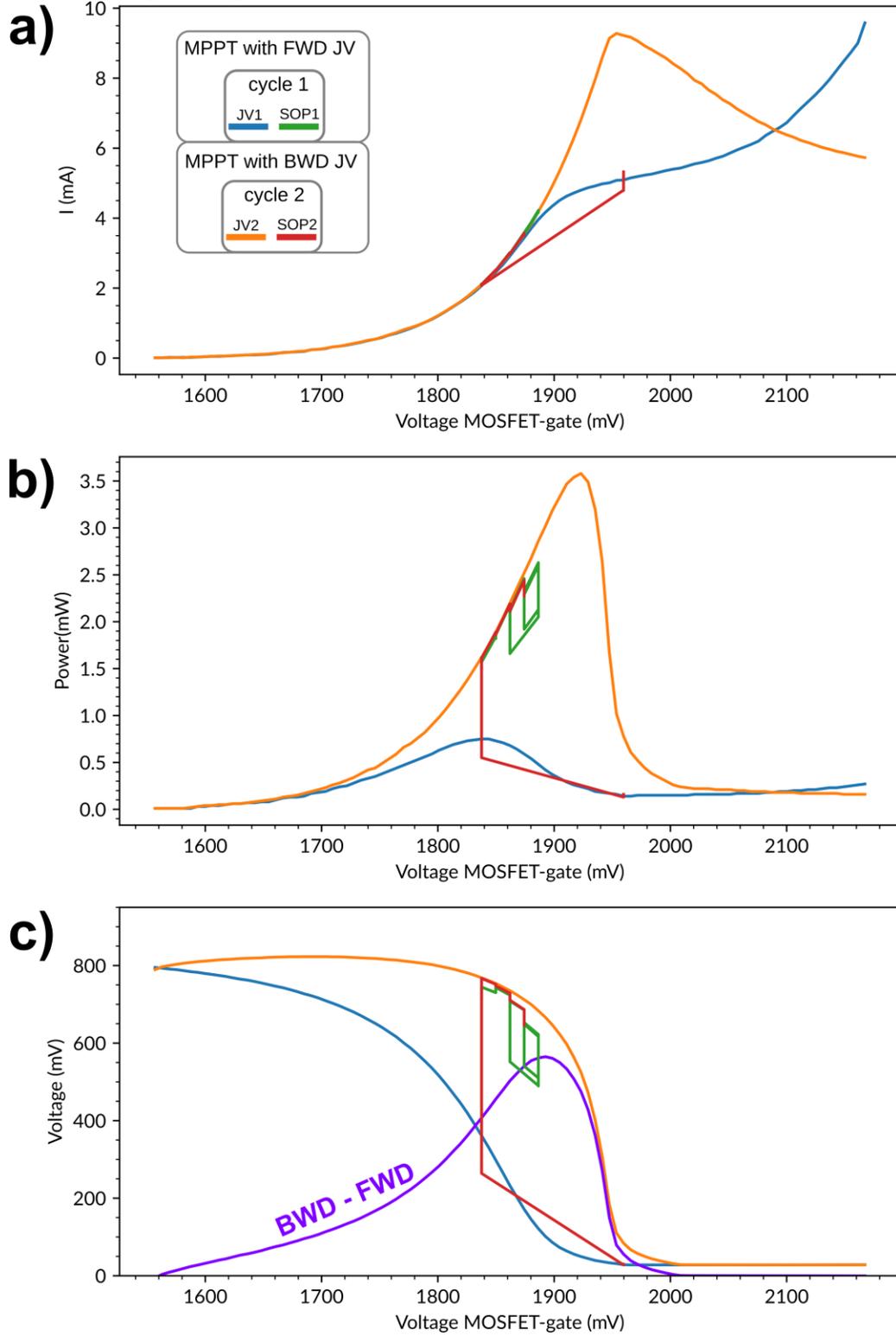

**Figure S7**. Transfer curves between $V_{GATE}$ and a) current, b) power and c) voltage of the cell during the manual operation of the MPP tracker. Colors follow the convention of Figure 8.

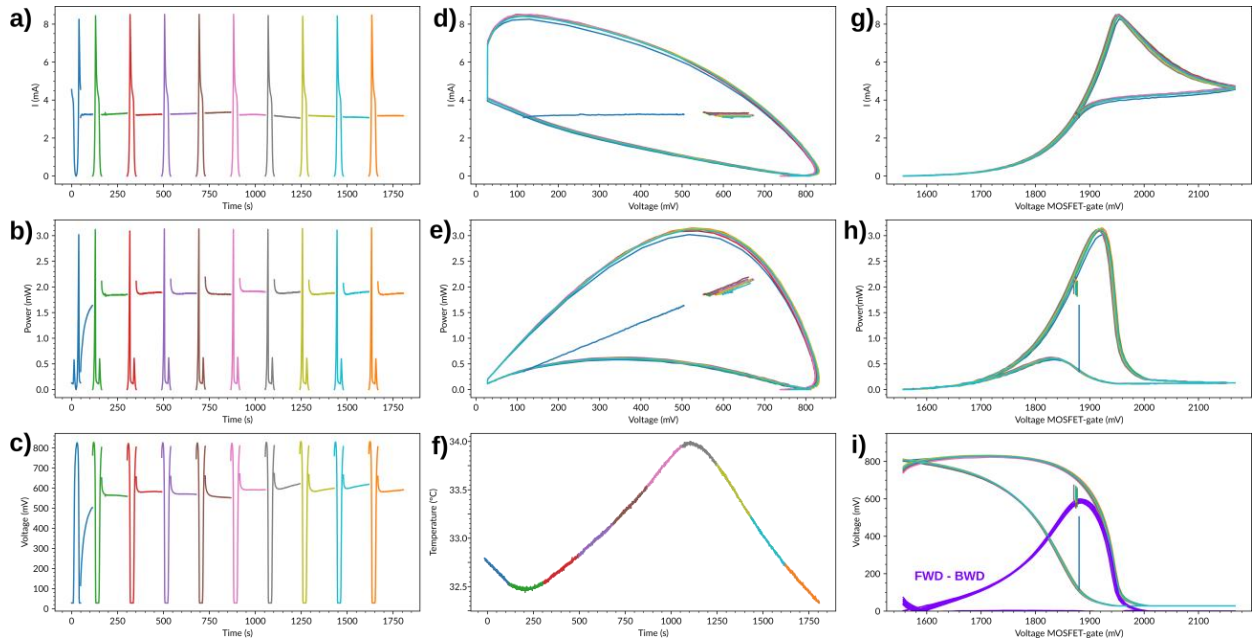

**Figure S8.** Ten automatic MPPT search cycles using the galvanostatic approach and power *vs* V$_{GATE}$ transfer curve. First cycles used FWD-BWD JV scan loops and the nine remaining used BWD-FWD JV scan loops for optimal V$_{MPP}$ determination. After that, the SOP stages have 120 s of duration. a) Instantaneous current, b) power and c) voltage during the MPPT sequence alternating JV loop and SOP stages under automatic MPP search. d) JV and e) Power-voltage curve output from the solar cell during JV scans and SOP stages. f) Solar cell temperature during the MPPT sequence. V$_{GATE}$ vs g) cell current, h) power output and i) voltage cell transfer functions.

# 4. Implementation of the algorithm for outdoor performance or variable irradiation.

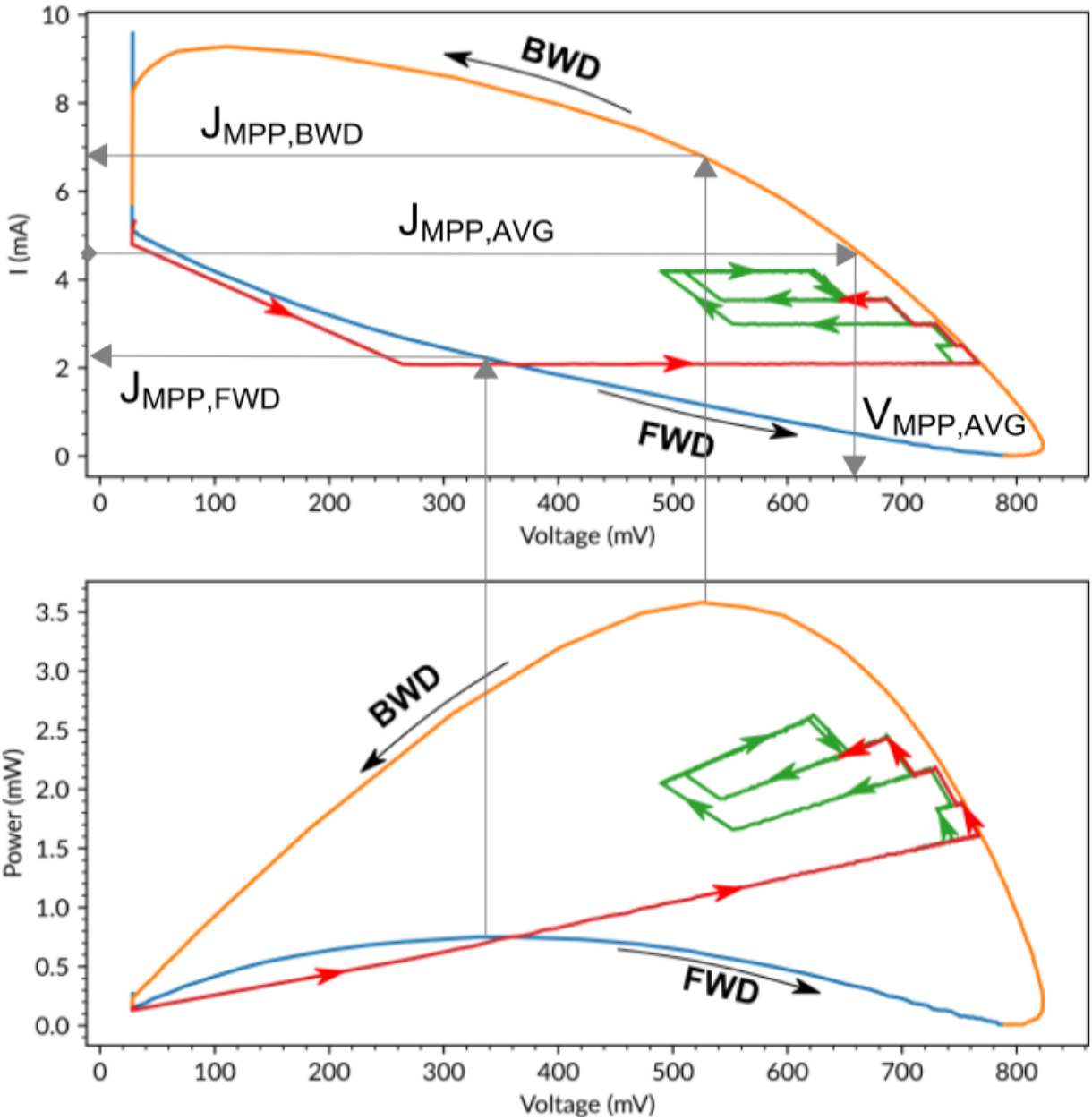

**Figure S9.** Graphical solution to the implementation of the galvanostatic algorithm in potentiostatic mode. Note that in absence of hysteresis this method results in the conventional approach.

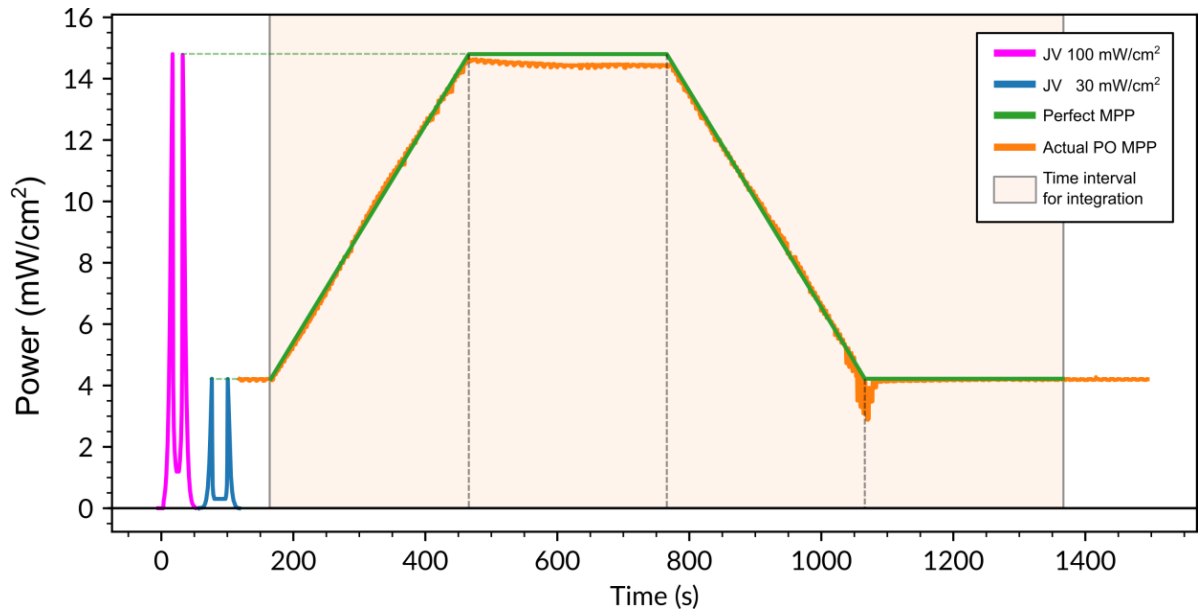

**Figure S10.** Instantaneous power output from a Si-cell guided by the small increment P&O MPP tracking test using one cycle 30/100 mW/cm$^2$ for 300 s ramping and dwelling stages. Pink (blue) traces are JV curves under 100 (30) mW/cm$^2$ irradiation from LED solar simulator. Green (red) trace denotes ideal MPP outputs considering ideal P$_{MPP}$ from BWD (FWD) JV curves at 30/100 mW/cm$^2$ irradiation.

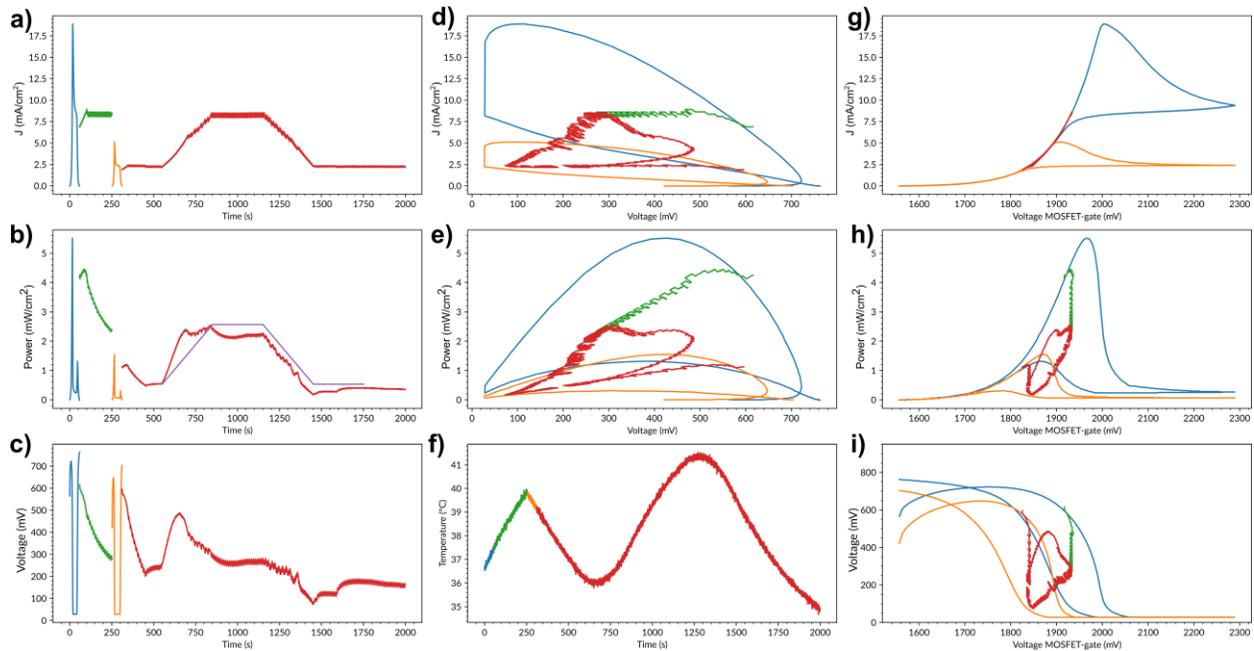

**Figure S11.** Conventional fixed increment P&O MPPT search cycles using the galvanostatic approach for the PSC under one cycle of variable illumination 30/100m W/cm$^2$ using 300 s of ramping and dwelling stages. In blue (orange) the 100 (30) mW/cm$^2$ BWD-FWD JV scan loop for optimal initial V$_{MPP}$ determination. a) Instantaneous current density, b) power output and c) cell voltage during the MPPT sequence. d) JV and e) Power-voltage curve output from the solar cell during the MPPT sequence. f) Solar cell temperature during the MPPT sequence. V$_{GATE}$ *vs* g) current density, h) power output and i) cell voltage transfer functions.

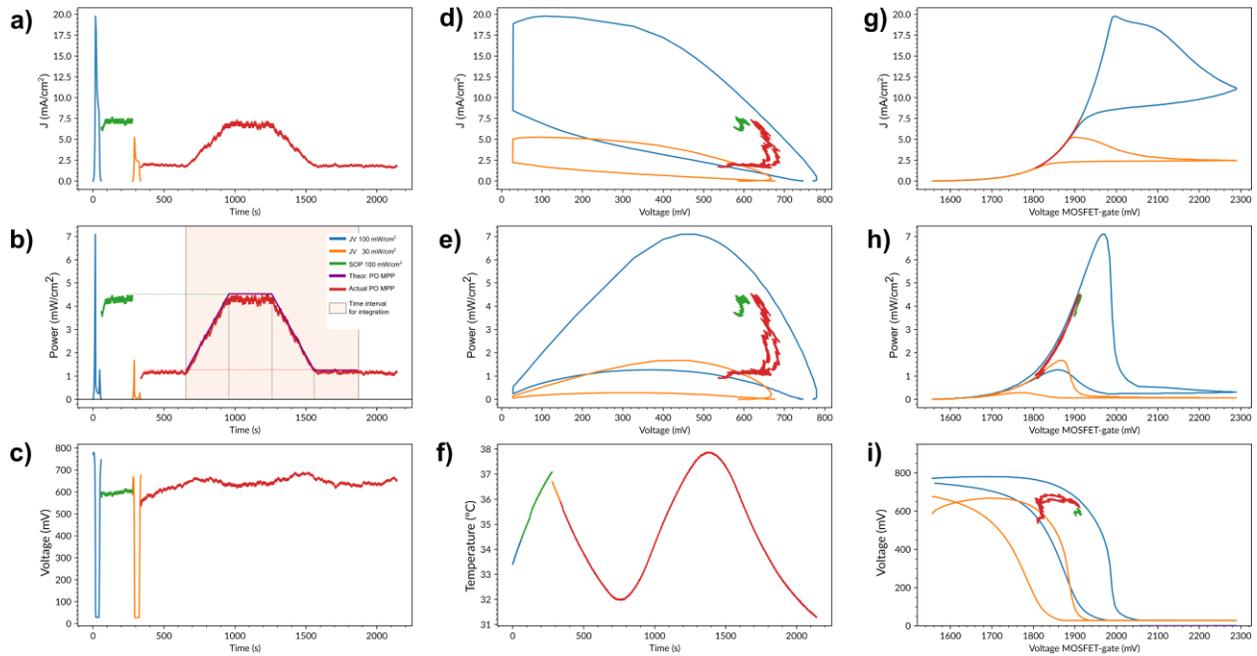

**Figure S12.** *k-feedback* fixed increment P&O MPPT search cycles using the galvanostatic approach for the PSC under one cycle of variable illumination 30/100 mW/cm$^2$ using 300 s of ramping and dwelling stages. In blue (orange) the 100 (30) mW/cm$^2$ BWD-FWD JV scan loop for optimal initial V$_{MPP}$ determination. a) Instantaneous density current, b) power output and c) cell voltage during the MPPT sequence. d) JV and e) power-voltage curve output from the solar cell during the MPPT sequence. f) Solar cell temperature during the MPPT sequence. V$_{GATE}$ *vs* g) current density, h) power output and i) cell voltage transfer functions.

# 5. Github repository

Code, circuit diagrams, bill of materials and gerber files for the shield PCB deposited in:

https://github.com/ej-jp/perovskino